\documentclass[%
nofootinbib,
 amsmath,amssymb,
 aps,
prc,
]{revtex4-2}

\usepackage{graphicx}
\usepackage{dcolumn}
\usepackage{bm}
\usepackage{hyperref}

\usepackage{color,xcolor,soul}

\DeclareMathOperator{\Div}{div}
\DeclareMathOperator{\Rot}{curl}

\renewcommand{\vec}{\mathbf}

\newcommand{\vp}{{\vec p}}

\newcommand{\vq}{{\vec q}}              
\newcommand{\vps}{{\vp '}}      
              
\renewcommand{\vr}{{\vec r}} 
\newcommand{\vrd}{{\pmb \rho}}

\newcommand{\vQ}{{\vec Q}}

\newcommand{\vVra}{{\vec V_{\alpha}}}   
   
\newcommand{\vVre}{{\vec V_{e}}}   
 
\newcommand{\vVrn}{{\vec V_{n}}}

\newcommand{\vj}{{\vec j}}       
\newcommand{\vA}{{\vec A}}       
\newcommand{\vB}{{\vec B}}

\newcommand{\vvg}{{\pmb v_\vp}}    
\newcommand{\vg}{{ v_\vp}}

\newcommand{\vv}{{\vec V_L}}     

\newcommand{\mN}{\mbox{$\mathcal N$}}

\newcommand{\uuq}{\mbox{$u_{\vq}^{(p)}$}}
\newcommand{\vvq}{\mbox{$v_{\vq}^{(p)}$}}
\newcommand{\uuqsq}{\mbox{$u_{\vq}^{(p)2}$}}
\newcommand{\vvqsq}{\mbox{$v_{\vq}^{(p)2}$}}

\begin{document}
\title{Neutron contribution to the force on a proton vortex in superconducting neutron-star matter}

\author{Oleg A.~Goglichidze}
\email{goglichidze@gmail.com}
\author{Mikhail E.~Gusakov}
\affiliation{Ioffe Institute, Politekhnicheskaya 26, 194021 St.~Petersburg, Russia}

\date{\today}

\begin{abstract}
    We investigate the forces acting on a proton vortex in superconducting 
    neutron-star matter composed of neutrons, protons, and electrons,   
    accounting for Fermi-liquid interactions in the neutron–proton subsystem.
    While the force arising from electron scattering by the vortex magnetic field is well known, we demonstrate that normal neutrons also exert a force on proton vortices due to Fermi-liquid coupling with superconducting protons.
    Using a kinetic approach based on Landau Fermi-liquid theory, we show that neutron quasiparticles scatter off a proton vortex through
    the spatially varying condensate momentum, giving rise to a longitudinal force proportional to the relative neutron–vortex velocity.
    In contrast to the electron contribution, this force has no transverse component and vanishes in the absence of neutron–proton Fermi-liquid interaction.
    Within a simple model, we analyze the main properties of this force and provide physically motivated estimates of its magnitude.
\end{abstract}

\maketitle

\section{Introduction}

    Studying neutron stars (NS), 
    the densest and most strongly magnetized objects known,
    requires a careful treatment of a wide range of physical phenomena, including nuclear interactions, superfluidity, superconductivity, and processes occurring in strong electromagnetic fields.
    Although the value of the proton pairing gap remains a topic of active theoretical research (see, e.g., Ref.~\cite{SedrakianClark2019}),
    it is widely believed that a substantial fraction of the NS core becomes superconducting once its temperature drops below a few times $10^9$~K.
    While the properties of the deeper core regions remain more uncertain, the outer layers are expected to enter the type-II superconducting state \cite{GlampedakisAnderssonSamuelsson2011}.
    In type-II superconductors, the magnetic field is confined within proton vortices (flux tubes).
    Therefore, understanding the forces acting on these vortices is required for accurate modeling of the NS magnetic field evolution, as well as for calculating the damping rates of various stellar oscillation modes.

    The forces acting on proton vortices (or similar objects)
    due to their motion relative to the surrounding matter
    have been the subject of numerous studies \cite{GlampedakisAnderssonSamuelsson2011,AlparLangerSauls1984,Jones1991b,Jones2006,Jones2009,AlfordSedrakian2010,GraberEtAl2015,BransgroveLevinBeloborodov2018,Gusakov2019,SourieChamel2020,GG25}.
    However, the analyses in most of these works were carried out under the assumption of vanishing temperature, with protons and neutrons treated as entirely superconducting/superfluid.
    Under such conditions, the normal (nonsuperfluid) component consists solely of leptons, which scatter off the vortex magnetic field.
    Therefore, the force acting on the vortex arises as a reaction to the Lorentz force exerted on the leptons.
    In fact, in the zero-temperature limit, the lepton contribution may be regarded as the only 
    velocity-dependent force acting on the proton vortex \cite{Gusakov2019,GG25}.

    What would change if we allow for the nucleonic thermal excitations?
    It is natural to expect that, if there exists a mechanism by which the thermal excitations can sense the presence of a vortex, an additional contribution to the total force may arise.
    To the best of our knowledge, Jones (2009) \cite{Jones2009} is one of the very few works that attempts to account for nucleonic thermal excitations in NS
    proton-vortex dynamics, considering interactions with quasiparticle excitations bound to the vortex core.
    Thus, the role of the normal nucleon component remains underexplored.

    It is well known from the theory of conventional superconductors that the energy of a Bogoliubov excitation depends explicitly on the condensate momentum $\vQ$ \cite{deGennes_book}.
    To linear order in $\vQ$, it takes the form
    \begin{equation}
        \label{eq:bog_ex}
        \varepsilon_\vp = \frac{\vp  \vQ}{m} + \sqrt{ \left(\epsilon_\vp - E_F \right)^2 + \Delta^2  },
    \end{equation}
    where $m$ is the (quasi)particle mass,
    $\epsilon_\vp$ is the (quasi)particle energy in the normal state, $E_F$ is the Fermi energy, and $\Delta$ is the superfluid energy gap.
    The spatially varying condensate momentum can be treated as a vector field which can scatter the Bogoliubov excitations. 
    This scattering gives rise to a force whose transverse component is often referred to as the Iordanskii force \cite{SoninBook}.

    In nucleonic matter, the dependence of the excitation energy on the condensate momentum is more involved.
    The neutron--proton subsystem forms a mixture in which Fermi-liquid effects play a significant role.
    These effects imply that the energy of a quasiparticle of a given species $\alpha$, say $\alpha = n$, depends on the distribution functions of quasiparticles of both species, $\alpha' = n, p$.
    In particular, one can expect that, in such a mixture, quasiparticles of both species interact with the condensate momentum associated with a proton vortex.
    In the present paper, we show that this interaction gives rise to a force exerted on a proton vortex by the neutron component, proportional to the incident flux of normal neutrons.

    The paper is organized as follows.
    In Sec.~\ref{sec:problem}, we formulate the problem and discuss the main assumptions.
    Sec.~\ref{sec:npe} describes the interaction of the different constituents of the mixture with a proton vortex.
    This section is divided into three subsections devoted, respectively, to protons (together with the electromagnetic field), electrons, and neutrons.
    In Sec.~\ref{sec:force}, we calculate the contributions of the different constituents to the total force.
    In Sec.~\ref{sec:dis}, we discuss the obtained results and
    present our conclusions.
    For clarity of presentation, most of the technical derivations are relegated to Appendices~\ref{sec:energy}--\ref{sec:ap:H2_contr}.

\section{Problem formulation}
\label{sec:problem}

    We consider a mixture consisting of neutrons, protons, and electrons. 
    The neutrons and protons are assumed to form a Fermi-liquid mixture.
    In the present paper, we consider a mixture at temperatures $T$ well below the proton critical temperature $T_{cp}$, which allows us to neglect proton thermal excitations.
    At the same time, we assume that $T>T_{cn}$, where $T_{cn}$ is the neutron critical temperature. Therefore, the neutron component remains nonsuperfluid.

    We assume the following hierarchy of length scales:
    \begin{equation}
        \label{eq:ineq}
        \hbar/p_{F\alpha} \ll \xi \ll \lambda \ll d_v \ll \ell_\alpha,
    \end{equation}   
    where $p_{F\alpha}$ are the Fermi momenta of particle species $\alpha$, $\xi$ is the proton coherence length, $\lambda$ is the London penetration depth, $d_v$ is the intervortex distance, and, finally, $\ell_\alpha$ are the characteristic mean free paths of particle species $\alpha$ in the absence of vortices.
    These inequalities allow us to substantially simplify the problem.
    Owing to the large mean free paths $\ell_\alpha$ compared with all other characteristic length scales, we can neglect collisions between (quasi)particles.
    More rigorous conditions on $\ell_\alpha$ that justify the ballistic regime for the considered problem are discussed in Secs.~\ref{sec:electrons} and \ref{sec:neutrons}.
    The large intervortex distance $d_v$ compared with the London penetration depth $\lambda$ justifies neglecting the influence of neighboring vortices, allowing us to consider an isolated proton vortex.

    Among the conditions \eqref{eq:ineq}, the least well-justified is that relating $\xi$ and $\lambda$. 
    These two quantities are typically of comparable magnitude in NS interiors \cite{GlampedakisAnderssonSamuelsson2011}.
    In the present paper, however, we adopt the simplifying assumption that the vortex core is treated as infinitely thin, while the London penetration depth remains finite.
    Although this assumption is not entirely realistic for NS interiors, it substantially simplifies the analysis and allows us to isolate the physical mechanisms underlying the forces under consideration.
    The limitations of this model are discussed in Sec.~\ref{sec:dis}.

    The vortex is assumed to be straight, enabling us to introduce cylindrical coordinates $(\rho, \phi, z)$ with the $z$ axis coinciding with the vortex axis and directed such that the integral \eqref{eq:circ} is positive, see below.
    Various particle species are allowed to flow through the vortex.
    The incident fluxes are taken to be uniform and stationary.    
    We further assume that the electric field vanishes far from the vortex.
    Consequently, the quasineutrality condition
    \begin{align}
        n_{p0}=n_{e0}
        \label{eq:quasi}
    \end{align}
    is satisfied for the unperturbed proton and electron densities, $n_{p0}$ and $n_{e0}$, respectively.

    The main goal of the present paper is to analyze the forces acting on the proton vortex induced by the incident particle fluxes. 
    To define these forces,
    we utilize a momentum conservation equation, which, for our system, has the following form (see Appendix~\ref{sec:mom_cons} and Refs.~\cite{Gusakov2019,GG25}):
    \begin{align} 
        \frac{\partial \mathfrak{P}^i}{\partial t}
	+ \frac{\partial \Pi^{ik}}{\partial r^k}
	=  \delta^{(2)}(\vrd)\, {F}_{\rm ext}^i,
        \label{eq:mom_eq_ap} 
    \end{align}
    where the total momentum density $\mathfrak{P}^i$ and the total stress tensor $\Pi^{ik}$ consist of contributions from superconducting $npe$ matter as well as from the electromagnetic field.
    The exact expressions for these quantities are given by Eqs.~\eqref{eq:total_mom} and \eqref{eq:total_stress}, respectively.
    The right-hand side of Eq.~\eqref{eq:mom_eq_ap} contains a term that we will refer to as ``external force''.
    This term is added to ensure the momentum conservation in the general case (as discussed shortly below).
    The two-dimensional delta function is explicitly written to emphasize that the external force is applied to the infinitely thin vortex core.

    If all the incident fluxes are independent of time, the problem is stationary. 
    In this case, the time derivative in Eq.~\eqref{eq:mom_eq_ap} is zero. 
    Let us choose a cylinder of unit length and radius $R$, whose axis coincides with the vortex axis.
    Integrating Eq.~\eqref{eq:mom_eq_ap} over the volume of this cylinder, we obtain a force balance equation of the following form:
    \begin{equation}
	\label{eq:force_bal}
	   \vec{F}_{\rm m \rightarrow v}  
	   +  \vec{F}_{\rm ext}  = 0.
    \end{equation}
    In this equation, the first term is defined as
    \begin{equation}
	\label{eq:F_mv_def}
	{F}_{\rm m \rightarrow v}^i 
             = - R \int\limits_0^{2 \pi} \Pi^{i\rho}  d\phi.
    \end{equation}
    In deriving $\vec{F}_{\rm m \rightarrow v}$, 
    the volume integral was converted into the surface integral using Gauss's theorem.
    Accordingly, the integration over cylindrical surface reduces to the integration over the azimuthal angle, multiplied by the radius $R$  and the unit length of the cylinder. 
    The quantity $\vec{F}_{\rm m \rightarrow v}$ describes the momentum flux entering the cylinder.
    It can be interpreted as the total force (per unit length) acting on the vortex from the surrounding matter.
    Therefore, the force given by Eq.~\eqref{eq:F_mv_def} is the subject of our analysis.

    The second term in the force balance equation \eqref{eq:force_bal} represents the external force applied to the vortex core, as previously stated. 
    This external force counteracts the force  $\vec{F}_{\rm m \rightarrow v}$.
    But why does such an external force appear in the force balance equation?
    Recall that the vortex is immersed in stationary incident fluxes and prescribed to remain at rest.
    However, in the general case, the total force $\vec{F}_{\rm m \rightarrow v}$, which includes the contributions from the various particle species, does not vanish unless the particle fluxes are fine tuned.
    Consequently, to maintain the stationarity, it must be counterbalanced by some 
    force, which, in our model,  is represented by the formally introduced term $\vec{F}_{\rm ext}$.
    In real NSs, such a force can be provided by buoyancy or tension effects \cite{Gusakov2019,dg17}, as well as by pinning to neutron vortices.
    A  more detailed discussion of Eq.~\eqref{eq:force_bal} can be found in Ref.~\cite{GG25}.

\section{Interaction of $npe$ mixture with a proton vortex}
\label{sec:npe}

    The stress tensor appearing in Eqs.~\eqref{eq:mom_eq_ap} and \eqref{eq:F_mv_def} includes contributions from protons, neutrons, electrons, and the electromagnetic field [see Eq.~\eqref{eq:total_stress}].
    Therefore, before evaluating the force, we must first determine the behavior of each constituent in the vicinity of the vortex. The present section is devoted to this task.

\subsection{Protons and the electromagnetic field}
\label{sec:protons}

    To describe the superconducting protons,
    we introduce the condensate momentum \cite{AronovEtAl1981}
    \begin{equation}
	   \label{eq:Q_def}
	   \vQ_p = \frac{\hbar}{2} \nabla \chi_p - \frac{e}{c} \vec A,
    \end{equation}
    and the non-equilibrium proton chemical potential \cite{AronovEtAl1981, Gusakov2010, GG23}
    \begin{equation}
	   \label{eq:mu_def}
	   \breve{\mu}_p = E_{Fp}-\frac{\hbar}{2} \frac{\partial \chi_p}{\partial t} - e \varphi,
    \end{equation}
    where $\chi_p$ is the order parameter phase, $\varphi$ and $ \vec A$ are the electrostatic  and vector potentials,  $e = |e|$ is the elementary charge, and by $E_{Fp}$ we denoted the proton Fermi energy calculated in the absence of the vortex as well as any currents.
    The vortex manifests itself through a nonzero circulation of the phase gradient:    
    \begin{equation}
	   \label{eq:circ}
	   \frac{\hbar}{2}\oint \nabla \chi_p d \vec{l} = m_p \varkappa,
    \end{equation}
    where the integral is taken along an arbitrary contour 
    encircling the $z$ axis once in the counterclockwise direction as seen from $+z$ and 
    \begin{equation}
        \label{eq:kappa_def}
	    \varkappa = \frac{2 \pi  \hbar}{ 2  m_p}
    \end{equation}
    is the circulation quantum.

    The vector potential can be represented as $\vA = \vA_{v} + \delta \vA$, where $\vA_{v}$ is the vector potential generated by the vortex while $\delta \vA$ is a small correction to $\vA_{v}$ arising  due to the nonzero particle flows streaming through the vortex. 
    The vortex vector potential can be represented as 
    \begin{equation}
	       \label{eq:A}
	       \vec{A}_v = \frac{c}{e}\frac{m_p \varkappa}{2 \pi \rho} \mathcal{P}(\rho) \vec{e}_\phi,
    \end{equation}
    where $\mathcal{P}(\rho) $ 
    is a function that will be specified below.
    The magnetic field, in turn, equals $\vB = \Rot \vA = \vB_{v} + \delta \vB$,
    where
    \begin{equation}
	       \label{eq:B_vortex}
	       \vec{B}_v
	       = \Rot \vec{A}_v = \frac{c}{e}\frac{m_p \varkappa}{2 \pi \rho} \mathcal{P}'(\rho)\vec{e}_z,
    \end{equation}
    $\mathcal{P}'(\rho)$ 
    denotes
    the derivative of the function $\mathcal{P}(\rho)$,
    and $\delta \vB = \Rot \delta \vA$.
    In Eqs.~\eqref{eq:A} and \eqref{eq:B_vortex}, $\vec{e}_\phi$ and $\vec{e}_z$ denote the unit vectors in the $\phi$ and $z$ directions.
    One can also calculate
    the magnetic flux enclosed within a circle of a radius $\rho$: 
    \begin{equation}
        \label{eq:Phi_B_def}
        \Phi_{B}(\rho) = \int\limits_{\rho} d^2\vrd_1 \ 
        \vec{e}_z  \vec{B}_v(\rho_1) = \Phi_0 \mathcal{P}(\rho),
    \end{equation}
    where $\Phi_0 = c m_p \varkappa / e$ is the total vortex magnetic flux.
    Hence, the function $\mathcal{P}(\rho)$ can be interpreted as the dimensionless magnetic flux enclosed by the
    circle
    of radius $\rho$.
    It varies from 0 at the vortex axis to 1 at distances much larger than $\lambda$.
    To determine the exact form of the function $\mathcal{P}(\rho)$, one must solve the microscopic quantum problem in conjunction with Maxwell’s equations.
    However, for the case of extreme type-II superconductivity ($\xi\ll\lambda$), it can be easily obtained and equals \cite{deGennes_book,Gusakov2019}
    \begin{equation}
        \label{eq:P_val}
	       \mathcal{P}(\rho) = 1 - \frac{\rho}{\lambda} K_1\left(\frac{\rho}{\lambda}\right).
    \end{equation}
    and, consequently, 
    \begin{equation}
	       \label{eq:dP_apr}
	       \mathcal{P}'(\rho)= \frac{\rho}{\lambda^2} K_0\left(\frac{\rho}{\lambda}\right),
    \end{equation}   
    where $K_n$ denotes the modified Bessel function of the second kind.

    The total condensate momentum \eqref{eq:Q_def} can be represented as
    \begin{equation}
	       \label{eq:Q_mod_ch}
	       \vQ_p
	       = \vQ_{p0} + \vQ_{pv} + \delta \vQ_p.
    \end{equation}
    Here, the first term, $\vQ_{p0}$,  is a constant vector describing the homogeneous flow measured far from the vortex.
    The second term,
    \begin{equation}
 	      \label{eq:Q_pv_def}
 	      \vQ_{pv} =  \frac{m_p \varkappa}{2 \pi \rho}\left[1 -  \mathcal{P}(\rho)\right] \vec{e}_\phi,
    \end{equation}
    represents the vortex flow.
    Finally, the third term of Eq.~\eqref{eq:Q_mod_ch} is the correction induced by the interaction of the incident fluxes with the vortex.
    A detailed analysis of this correction can be found in Ref.~\cite{GG25} for the case of vanishing temperature.
    For the purposes of the present study, the exact form of $\delta \vQ_p$ is not required.
    However, in the  following calculations, we will use the fact  that this correction is linear in the incident flux and decays as $\rho^{-1}$ for $\rho \gg \lambda$.

    The subsequent calculations require the curl of the condensate momentum, which equals
    \begin{equation}
	\Rot\vQ_p  
        = \frac{m_p \varkappa}{2 \pi \rho}\delta(\rho) \vec{e}_z- \frac{e}{c} \vec{B}
        = \Rot \vQ_{pv} - \frac{e}{c} \delta \vec{B},
	\label{eq:rotQ_ch}
    \end{equation}
    where 
    \begin{equation}
        \label{eq:rotQpv}
         \Rot \vQ_{pv} = -\frac{m_p \varkappa}{2 \pi \rho}  \tilde{\mathcal{P}}'(\rho) \vec{e}_z,
    \end{equation}
    and we introduced the function 
    \begin{equation}
    	\label{eq:dPt_def}
    	\tilde{\mathcal{P}}'(\rho) = {\mathcal{P}}'(\rho) - \delta(\rho)
    \end{equation}
    with $\delta(\rho)$ denoting the Dirac delta function.
    The appearance of the delta function in Eq.~\eqref{eq:dPt_def} requires some clarification. 
    It formally arises from taking the $\Rot$ of the superfluid phase gradient. 
    It is well known from vector analysis that taking the $\Rot$ of a gradient of any well-defined, single-valued function yields identically zero.
    However, the phase $\chi_p$ and, consequently, the condensate momentum $\vQ_p$ are not defined (singular) at the vortex axis. 
    This singularity arises due to the nonzero circulation of the phase gradient.
    Using the identity~\eqref{eq:circ}, we define the formal construction $\Rot \nabla \chi_p$ at $\rho = 0$ by means of Stokes' theorem:
    \begin{equation}
	\frac{\hbar}{2} \int \Rot\nabla  \chi_p \vec{n} d{S} = \frac{\hbar}{2} \oint \nabla  \chi_pd\vec{l} =m_p \varkappa,
    \end{equation}
    where the left-hand integration is performed over an arbitrary (oriented) surface punctured by the vortex ($\vec{n}$ is the unit vector normal to the surface), while  the right-hand integration is taken over the oriented boundary of this surface.
    Thus, for a straight vortex, we have:  $\Rot \nabla \chi_p=2 \pi \delta^{(2)}({\vrd})\, \vec{e}_z$.
    To arrive at Eq.~\eqref{eq:rotQ_ch}, we applied the identity:
    \begin{equation}
        \label{eq:deltas_rel}
        \delta^{(2)}(\pmb \rho)=\frac{\delta(\rho)}{2 \pi \rho}.
    \end{equation}
    By analogy with the vortex magnetic field $\vB_v$, we define
    the flux of the vector 
    $\Rot\vQ_{pv}$
    enclosed within the circle of a radius $\rho$:
    \begin{equation}
        \label{eq:Phi_Q_def}
        \Phi_{Q}(\rho) = \frac{c}{e} \int\limits_{\rho} d^2\vrd_1 \ \vec{e}_z \Rot 
        \vec{Q}_{pv}(\rho_1)  
        = \frac{c}{e} \oint \limits_{\rho} d\vec{l}\,  
        \vec{Q}_{pv}
        =   \Phi_0 \left[1 -  \mathcal{P}(\rho)\right].
    \end{equation}
    The factor $c/e$ is used to make the dimension of $\Phi_{Q}$ consistent with that of $\Phi_{B}$.
    Since the function $\mathcal{P}(\rho)$ tends to unity at large distances,
    the total flux $\Phi_{Q}$ equals zero.
    The vanishing of the total flux $\Phi_{Q}$ can be regarded as a consequence of the Meissner effect, resulting in the exponential suppression of the vortex circular current.

    We assume that the proton incident flux is small, specifically, 
    \begin{equation}
        \label{eq:p_ineq}
        \iota_p = \frac{Q_{p0}}{p_{Fp}} \ll 1.
    \end{equation}
    Separately, we assume the smallness of the following ratio:
    \begin{equation}
        \label{eq:Qpv_ineq}    
        \frac{Q_{pv}}{p_{Fp}} \ll 1.
    \end{equation}  
    The latter condition is clearly violated in the vicinity of the vortex axis since
    $Q_{pv}$ diverges as $\rho \rightarrow 0$.
    As it was stated, we adopt a model where the vortex core is approximated by an infinitely thin line.
    However, in reality, the vortex core has a finite radius of the order of the coherence length $\xi$
    given by the following formula:
    \begin{equation}
	   \label{eq:xi_def}
    	\xi = \frac{\hbar p_{Fp}}{\pi \Delta^{(p)} m_p^*},
    \end{equation} 
    where $\Delta^{(p)}$ is the proton energy gap and 
    $m_p^*$ is the proton effective mass.
    Thus, in the region outside the core (where our model is valid), we can 
    estimate:
    \begin{equation}
        \label{eq:Qpv_est_n}
	       \frac{{Q}_{pv}}{p_{Fp}} \lesssim \frac{m_p \varkappa}{2\pi \xi p_{Fp}}.
    \end{equation}
    Introducing the small parameter 
    \begin{equation}
	\label{eq:eps_p_def}
	       \epsilon =     \frac{m_p^*\Delta^{(p)}}{ p_{Fp}^2},
    \end{equation}
    and making use of definitions    
    \eqref{eq:xi_def} and \eqref{eq:kappa_def}, from Eq.~\eqref{eq:Qpv_est_n}, we get:
    \begin{equation}
        \label{eq:Qpv_est_n2}
	\frac{{Q}_{pv}}{p_{Fp}} \lesssim 
   \frac{\pi}{2} \epsilon.    
    \end{equation}
    In what follows, we expand the quantities of interest as a power series in the small parameter $\epsilon$. 
    In particular, the quantities $\delta \vA$, $\delta \vB$, and $\delta \vQ_p$ can be expanded in $\epsilon$ starting from the linear term.

    Strictly speaking, there is another dimensionless parameter in the problem:
    \begin{equation}
        \label{eq:another_small_p}
        \frac{Q_{p} p_{Fp}}{m_p \Delta^{(p)}},        
    \end{equation}
    which cannot be considered small everywhere. 
    Finite values of this parameter lead to the existence of bounded Bogoliubov excitations even in the limit of vanishing temperature.
    This complicates the expression for the proton current density, reduces the proton gap compared with its unperturbed value, and eventually results in the breakdown of the superconductivity  \cite{deGennes_book}. 
    In our model, we include the region where these effects become substantial in the vortex core.
    Thus, from the physical point of view, we effectively consider only the region $\rho > \xi$.

    In the stationary case, the chemical potential $\breve{\mu}_p$ is related to the electric field via the following equation [cf.~Eq.~\eqref{eq:superfluid_eq_ap}]:
   \begin{equation}
	   \label{eq:sf_eq}
	   \nabla  \breve{\mu}_p  = - e \nabla  \varphi =   e   \vec E.
    \end{equation}
    If there are no incident fluxes, the electric field is zero \cite{Gusakov2019} and, therefore, the function $\breve{\mu}_p$ remains unperturbed.
    The same is true if there is no vortex.
    Thus, accounting for the definition \eqref{eq:mu_def}, we can represent $\breve{\mu}_p$ as
    \begin{equation}
        \label{eq:mup_repr}
        \breve{\mu}_p = E_{Fp} + \delta \breve{\mu}_p,
    \end{equation}
    where $\delta \breve{\mu}_p$ is a small correction caused by the interaction of the vortex with the incident flux.
    The electric field and the correction $\delta \breve{\mu}_p$, in principle, should be determined self-consistently.
    However, as we will see, the exact form of these quantities is not required for the calculation of the force.

\subsection{Electrons}
\label{sec:electrons}

    The problem of electron scattering off a proton vortex has been previously studied in Refs.~\cite{Gusakov2019,GG25}.
    In this subsection, we provide a brief summary of the key results of those works.

    The electrons are treated as an ideal relativistic gas.
    Since the typical electron wavelength is much smaller than the other length scales, their behavior can be described using a distribution function $\mN_{\vp}^{(e)}$ satisfying  the Boltzmann-Vlasov equation:
    \begin{equation}
	\label{eq:kin_eq_e}
		\vvg \frac{\partial \mN_{\vp}^{ (e)}}{\partial \vr} 
		- e \left( \vec{E} +\frac{1}{c} \vvg \times \vec{B} \right) \frac{\partial \mN_\vp^{ (e)}}{\partial \vp} =  I^{ (e)},
    \end{equation}
    where $\vvg = \vp c^2 / \varepsilon_\vp^{(e)}$ and $ \varepsilon_\vp^{(e)} = \sqrt{m_e^2 c^4 + p^2 c^2}$.

    We assume that collisions between (quasi)particles are negligible; hence, the collision integral $I^{(e)}$ in Eq.~\eqref{eq:kin_eq_e} can be neglected.
    We will state the condition under which this approximation is justified at the end of this subsection.

    We further assume that, in the absence of the vortex, the electron distribution function reduces to the standard Fermi distribution.
    Accordingly, in the presence of the vortex, we seek the electron distribution function in the following form:    
    \begin{equation}
	\label{eq:Ne_ans}
	   \mN_{\vp}^{ (e)} = f_F\left( \varepsilon_\vp^{(e)} - E_{Fe}  - e \varphi - \vp  \vVre +  g_e \right),
    \end{equation}
    where 
    \begin{equation}
	   f_F(x) = \frac{1}{e^{x/T} +1 }
    \end{equation}
    is the Fermi distribution function,
    $E_{Fe}$ is the electron (unperturbed) Fermi energy,
    $\varphi$ is the electrostatic potential,
    $\vVre$ is the velocity of the electron incident 
    flux,
    and $g_e$ is a function of $\vr$ and $\vp$ to be determined.
    The argument of the function in Eq.~\eqref{eq:Ne_ans} is written to linear order in $\vVre$.  
    Since we will evaluate the force acting on the proton vortex in the linear approximation with respect to $\vVre$, this level of accuracy in Eq.~\eqref{eq:Ne_ans} is sufficient.
    The function $g_e$
    describes the correction
    due to the presence of the vortex. 
    Note that the electric field in the considered problem is induced solely by the interaction between the vortex and the incident charged fluxes.
    Therefore, the term $-e \varphi$
    should, strictly speaking, be regarded as part of the function $g_e$.
    However, by explicitly isolating this contribution, the electric interaction can be fully taken into account within the accepted level of accuracy and, consequently, the function $g_e$ in Eq.~\eqref{eq:Ne_ans} describes only the magnetic interaction with the vortex [see Eq.~\eqref{eq:dpe_eq_tmp2} below]. 
    Since a particle flow directed along the vortex cannot generate a force via the mechanisms considered in the present paper,
    we assume, without loss of generality, that $\vVre \perp \vec{e}_z$.

    It is instructive to examine small parameters governing the electron scattering problem.
    We assume that the velocity of the incident electron flux is small, specifically, 
    \begin{equation}
	       \label{eq:e_ineq}
	       \iota_e = \frac{V_{e}}{ v_{Fe} } \ll 1,
    \end{equation}
    where $v_{Fe}$ is the electron Fermi velocity. 
    As was already noted, the electric field $\vec{E}$ and the magnetic field correction $\delta \vB$ are generated by the interaction of the incident fluxes with the vortex. 
    Consequently, these quantities are at least linear in the parameter $\iota_e$ or in the analogous small parameters characterizing  the other incident fluxes [see Eqs.~\eqref{eq:p_ineq} and \eqref{eq:Vn_est}]

    Besides 
    these ratios,
    Eq.~\eqref{eq:kin_eq_e}    
    contains another small parameter,
    \begin{equation}
        \tilde{\epsilon} =  \frac{\hbar}{p_{Fe} \lambda} \sim \frac{e}{c} B_v \frac{\lambda}{p_{Fe}} \ll 1,
    \end{equation}
    where $B_v$ is estimated using Eq.~\eqref{eq:B_vortex}.%
    %
    \footnote{The smallness of the parameter $\tilde{\epsilon}$ means that the electrons pass through the vortex along  trajectories that are almost rectilinear.}
    %
    Utilizing Eq.\ \eqref{eq:xi_def} and bearing in mind that $p_{Fe} = p_{Fp}$, we can establish a relationship between this small parameter and 
    the parameter $\epsilon$ introduced in Sec.~\ref{sec:protons} [see Eq.~\eqref{eq:eps_p_def}]:
    %
    \begin{equation}
        \tilde{\epsilon} = \pi \frac{\xi}{\lambda} \epsilon.
    \end{equation}
    Since $\xi < \lambda$, the smallness of $\epsilon$ ensures that $\tilde{\epsilon}$ is also small.
    The same parameter will appear a third time when considering neutrons.
    This allows us to treat $\epsilon$ as a universal small parameter governing the interaction of all the mixture constituents with the vortex.

    Substituting the ansatz \eqref{eq:Ne_ans} into Eq.~\eqref{eq:kin_eq_e} and  linearizing the result  with respect to the incident fluxes, we obtain the following equation for the function $g_e$:
    \begin{equation}
	       \label{eq:dpe_eq_tmp2}
	        \vp \frac{\partial  g_e}{\partial \vr} 
	       - \frac{e}{c}  (\vp \times \vec{B}_v)  \frac{\partial  g_e }{\partial \vp} 
	       = - \frac{e}{c} \vVre  \left(\vp\times \vec{B}_v \right).
    \end{equation}
    This equation can be integrated along the characteristics. 
    We choose the boundary condition so that the function $g_e$ vanishes far upstream of the scattering region.
    At the distances $\rho \gg \lambda$, the solution can be represented as
    \begin{equation}
        	\label{eq:dpe_ld}
        	g_{e} =  \left\{  - \left[\vp \left( \vec{e}_z \times \vVre \right) \right] \, \sigma_{\perp}^{(e)} +\left( \vp_\perp \vVre \right) \, \sigma_{||}^{(e)} \right\} \frac{\delta(\phi-\phi_p)}{\rho},
    \end{equation}
    where we introduced, respectively, the transverse and transport cross sections:
    \begin{align}
    	\label{eq:sigma_e_perp}
    	\sigma_{\perp}^{(e)}& = \frac{m_p \varkappa }{p_\perp} \, a_e,
    	\\
    	\label{eq:sigma_e_par}
    	\sigma_{||}^{(e)}  &=  
         \frac{(m_p \varkappa)^2 }{p_\perp^2} \, L_e^{-1},
    \end{align}
    and the following notations:
    \begin{equation}
    	\label{eq:ae_def}
    	a_e = \int\limits_{-\infty}^{\infty} dx \int \limits_{-\infty}^{\infty} dy\frac{\mathcal{P}'(\sqrt{x^2+y^2})}{2\pi \sqrt{x^2+y^2}},     
    \end{equation}
    \begin{equation}
    	\label{eq:Le_def}
    	L_e^{-1} = \frac{1}{2}\int\limits_{-\infty}^{\infty} dx \left( \int \limits_{-\infty}^{\infty} dy\frac{\mathcal{P}'(\sqrt{x^2+y^2})}{2\pi \sqrt{x^2+y^2}}   \right)^2.
    \end{equation}
    In these expressions, $\vp_\perp$ is the component of the vector $\vp$ perpendicular to $\vec{e}_z$, 
    $\varphi_p$ is its azimuthal angle.
    The double integral in Eq.~\eqref{eq:ae_def} is, in fact, 
    an integration over the entire plane perpendicular to the vector $\vec{e}_z$.
    Therefore, the coefficient $a_e$ 
    can be represented as 
    \begin{equation}
        \label{eq:ae_val}
        a_e = \lim_{\rho \rightarrow \infty} \frac{\Phi_B(\rho)}{\Phi_0} = 1,
    \end{equation}
    where the magnetic flux $\Phi_B(\rho)$ is given by Eq.~\eqref{eq:Phi_B_def}.    
    For the function $\mathcal{P}'$ given by Eq.~\eqref{eq:dP_apr}, the parameter $L_e^{-1}$ can be evaluated as well.
    It equals \cite{Gusakov2019,GG25}
    \begin{equation}
        \label{eq:Le_val}
        L_e^{-1} = \frac{1}{8\lambda}.
    \end{equation}

    A solution to Eq.~\eqref{eq:dpe_eq_tmp2} valid at arbitrary distance from the vortex axis can be found in Ref.~\cite{GG25} where Eq.~\eqref{eq:dpe_ld} is obtained as a limiting case. 
    A derivation of the solution \eqref{eq:dpe_ld} can also be found in Ref.~\cite{Gusakov2019}.
    Similar solutions for other superfluid/superconducting systems were obtained, e.g., in Refs.~\cite{Sonin1975,AronovEtAl1981}.
    Note that we restrict our analysis to terms up to second order in the small parameter $\epsilon$.%
    \footnote{One can say that  the expansion in small parameter $\epsilon$ is equivalent to an expansion in the dimensionful parameter $m_p\varkappa$.}
    In particular, $\sigma_{\perp}^{(e)} \sim \epsilon$, while $\sigma_{||}^{(e)} \sim \epsilon^2$.

    It remains to obtain a condition ensuring the validity of the ballistic approximation. 
    The collision integral in Eq.~\eqref{eq:kin_eq_e} can be estimated as
    \begin{equation}
         I^{ (e)}   \sim \frac{\vg}{\ell_e} \delta   \mN_{\vp}^{ (e)},
    \end{equation}
    where $ \delta \mN_{\vp}^{ (e)}$ is the departure of the function $\mN_{\vp}^{ (e)}$ from the equilibrium distribution function. 
    In our case, to the linear accuracy in the incident fluxes,
    \begin{equation}
        I^{ (e)}  \sim \frac{\vg}{\ell_e} \frac{\partial \mN_{\vp}^{ (e)} }{\partial \varepsilon_\vp^{(e)}} \left( g_e - \vp \vVre 
        \right).
    \end{equation}
    The collisions can be neglected when the collision integral is much less than the force term in Eq.~\eqref{eq:kin_eq_e}:
    \begin{equation}
       \frac{p \vg}{\ell_e} \ll \frac{e}{c} \vg B_v.
    \end{equation}
    Substituting Eqs.~\eqref{eq:kappa_def}, \eqref{eq:B_vortex},
    and noting that 
    the considered region has a characteristic length scale $\rho \sim \lambda$,
    we arrive at the following 
    condition \cite{Kopnin1995}:
    \begin{equation}
        \label{eq:le_cond}
        \ell_e \gg \frac{p_{Fe} \lambda^2}{\hbar}.
    \end{equation}
    This inequality is well satisfied under typical NS conditions.

\subsection{Neutrons}
\label{sec:neutrons}

    Let us finally consider the neutron constituent. 
    To describe it,
    we employ the Landau's theory of Fermi liquids \cite{PinesNozieres}.
    Being electrically neutral, 
    the neutron quasiparticles are unaffected by the vortex magnetic field.%
    \footnote{The interaction with the magnetic field via the neutron magnetic moment is small and can be neglected.}
    However, they can still sense the presence of the vortex through Fermi-liquid interaction: the neutron quasiparticle energy $\varepsilon_\vp^{(n)}$ depends on the proton distribution function $\mN_{\vp}^{(p)}$ [see Eq.~\eqref{eq:qp_energy_def}], which, in turn, depends on $\vQ_p$.

    The effect of a proton vortex on the neutron quasiparticle distribution function $\mN_{\vp}^{(n)}$ can be studied by solving the stationary Landau-Boltzmann kinetic equation
    \begin{equation}
    	\label{eq:kin_eq_n}
    	\frac{\partial \varepsilon_\vp^{(n)}}{ \partial \vp} \frac{\partial \mN_{\vp}^{(n)} }{ \partial \vr} 
    	- \frac{\partial \varepsilon_\vp^{(n)} }{ \partial \vr} \frac{\partial \mN_{\vp}^{(n)} }{ \partial \vp}	 = I_n.
    \end{equation}
    The right-hand side of this equation formally contains the collision integral $I_n$.
    However, similar to the case of electrons, 
    this term can be neglected
    (see the condition at the end of the present subsection).
    By analogy with the electron case, we seek the neutron distribution function in the following form:
    \begin{equation}
        \label{eq:Nn_ans_tmp}
        	 \mN_{\vp}^{(n)} = f_F\left( \varepsilon_\vp^{(n)} - E_{Fn}  - \vp  \vVrn +  g_n \right),
    \end{equation}    
    where $E_{Fn}$ is the unperturbed neutron Fermi energy
    and $g_n$ is a function of $\vr$ and $\vp$ to be found.
    The constant vector $\vVrn$ describes the neutron incident flux.
    Similar to the electron scattering problem, we assume that $\vVrn \perp \vec{e}_z$ and introduce the small parameter
    \begin{equation}
        \label{eq:Vn_est}
        \iota_n = \frac{m_n^* \vVrn}{p_{Fn}} \ll 1,
    \end{equation}
    where $m_n^*$ is the neutron effective mass defined via Eq.~\eqref{eq:mn_def}, see below. 
    Therefore, as in the case of electrons, we keep terms only to linear order in the velocity $\vVrn$ in the argument of the function \eqref{eq:Nn_ans_tmp}.

    As was already stated, 
    the neutron quasiparticle energy $\varepsilon_\vp^{(n)}$ depends on the neutron and proton distribution functions. 
    Therefore, strictly speaking,  Eq.~\eqref{eq:Nn_ans_tmp} together with the proton distribution function 
    should be considered as a system of implicit equations.
    However, taking into account  the smallness of the parameters defined in Eqs.~\eqref{eq:p_ineq}, \eqref{eq:eps_p_def}, \eqref{eq:e_ineq}, and \eqref{eq:Vn_est},  one can derive explicit expressions with the accuracy adopted in the present study.  
    Let us consider the neutron quasiparticle energy.
    In the presence of neutron and proton currents, it can generally be expanded as a power series in the parameter $\iota_n$ and the ratio $\vQ_p/p_{Fp}$:
    \begin{equation}
    	\label{eq:en_ser}
    	\varepsilon_\vp^{(n)}  = \varepsilon_{\vp,0}^{(n)} + \Delta H_{\vp,1}^{(n)} + \Delta H_{\vp,2}^{(n)} + ... 
    \end{equation}
    Note that, in Eq.~\eqref{eq:en_ser} we treat the ratio $\vQ_p/p_{Fp}$ as a single, unified small parameter.
    However, this expansion can, in principle, be rearranged as a double expansion in two separate small parameters given by Eqs.~\eqref{eq:p_ineq} and \eqref{eq:Qpv_ineq} [or Eq.~\eqref{eq:eps_p_def}].
    In Eq.~\eqref{eq:en_ser}, the unperturbed energy $\varepsilon_{\vp,0}^{(n)}$ depends only on the magnitude of the vector $\vp$, whereas the first order correction generally  has the following form \cite{GG23}:
    \begin{equation}
        \label{eq:DH1_def}   
        \Delta H_{\vp,1}^{(n)}  = \gamma_{np} \, \vp \vQ_p + K_{nn} \, \vp \vVrn,
    \end{equation}
    where the coefficients $\gamma_{np}$ and $K_{nn}$ are functions of Landau parameters and the effective masses.
    The second-order correction $\Delta H_{\vp,2}^{(n)}$ has a more complex structure. 
    In particular, it
    includes the contribution from the function $g_n$, which describes the scattered neutron quasiparticles. 
    Indeed, this function is proportional to the incident neutron flux $\sim \iota_n \sim \vVrn$ and at least linear in the parameter $\epsilon \sim \vQ_{pv}$ governing the interaction with the vortex [see the obtained solution below; cf.~Eq.~\eqref{eq:dpe_ld}]. 
    Therefore, the contribution from the function $g_n$ is at least quadratically small with respect to the expansion \eqref{eq:en_ser}.
    Fortunately, the exact form of the correction $\Delta H_{\vp,2}^{(n)}$ will not be required.

    As we will see, it is convenient to introduce a new momentum variable
    \begin{equation}
	       \label{eq:p_q_change}
        	\vq  = \vp + \gamma_{np} m_n^* \vQ_p,
    \end{equation}
    where  the neutron effective mass  $m_n^*$ is defined near the Fermi surface via the following equation:
    \begin{equation}
        \label{eq:mn_def}
        	 \frac{\partial \varepsilon_{\vp,0}^{(n)}}{ \partial \vp}  = \frac{\vp}{m_n^*}.
    \end{equation}
    In the terms of the new momentum variable, the neutron energy \eqref{eq:en_ser} becomes
    \begin{equation}
    	\label{eq:en_ser2}
	    \varepsilon_{\vq - \gamma_{np} m_n^* \vQ_p}^{(n)}  = \varepsilon_{\vq,0}^{(n)} + K_{nn} \, \vq \vVrn + \Delta \tilde{H}_{\vq,2}^{(n)} + ... ,
    \end{equation}
    where 
    $\Delta \tilde{H}_{\vq,2}^{(n)}$ is the new second order correction defined with respect to the momentum variable $\vq$.
    Under the change of variables from $(\vr,\, \vp)$ to $(\vr,\, \vq)$, Eq.~\eqref{eq:kin_eq_n}
    takes the form
    \begin{equation}
    	\label{eq:kin_eq_n2}
    	\pmb{v}_{gn} \frac{\partial \mN_{\vq - \gamma_{np} m_n^* \vQ_p}^{(n)} }{ \partial \vr} 
    	- \left[ \frac{\partial \varepsilon_{\vq - \gamma_{np} m_n^* \vQ_p}^{(n)} }{ \partial \vr}  +     \pmb{v}_{gn} \times \Rot \left( m_n^* \gamma_{np} \vQ_p \right)    \right] \frac{\partial \mN_{\vq - \gamma_{np} m_n^* \vQ_p}^{(n)} }{ \partial \vq} = 0,
    \end{equation}  
    while  the 
    distribution function 
    \eqref{eq:Nn_ans_tmp}
    becomes  
    \begin{equation}
    	\label{eq:Nn_ans}
	    \mN_{\vq - \gamma_{np} m_n^* \vQ_p}^{(n)} = f_F\left( \varepsilon_{\vq - \gamma_{np} m_n^* \vQ_p}^{(n)} - E_{Fn}  - \vq  \vVrn +  \tilde{g}_n \right),
    \end{equation}
    where 
    \begin{equation}
        \label{eq:g_n_tilde_def}
        \tilde{g}_n =  g_n  + \gamma_{np} m_n^* \vVrn \vQ_p.     
    \end{equation}
    In Eq.~\eqref{eq:kin_eq_n2}, we introduced the neutron group velocity
    \begin{equation}
        \label{eq:vgn_def}
	   \pmb{v}_{gn} = \frac{\partial \varepsilon_{\vq -  m_n^* \gamma_{np} \vQ_p}^{(n)}}{ \partial \vq}  = \frac{\vq}{m_n^*} +  K_{nn} \vVrn + \frac{\partial \Delta \tilde{H}_{\vq,2}^{(n)} }{ \partial \vq} + ...
    \end{equation}
    One can see that 
    the introduction of the new momentum variable $\vq$ offers two advantages.
    First, it allows to exclude the linear term $\propto \vQ_p$ (and, consequently,  $\propto \vQ_{pv}$) from the neutron group velocity \eqref{eq:vgn_def}.
    Second,  
    Eq.~\eqref{eq:kin_eq_n2},     
    written in the variables $(\vr,\, \vq)$, 
    closely resembles  
    Eq.~\eqref{eq:kin_eq_e}
    with $  \Rot \left( m_n^* \gamma_{np} \vQ_{p} \right) $ playing the role of a magnetic field.%
    \footnote{The factor $m_n^* \gamma_{np}$  is assumed to be constant outside the vortex core. 
    However, placing it under the $\Rot$ operator makes equations more concise and helps to clarify some subsequent reasoning.
    }
    Note, however, that Eq.~\eqref{eq:kin_eq_n2} contains the gradient of the neutron energy $\varepsilon_\vp^{(n)}$, which cannot be fully matched with the electric field that presents in Eq.~\eqref{eq:kin_eq_e}. 
    Indeed, unlike the electric field, 
    this gradient
    (a) generally depends on $\vq$, and (b) does not vanish even in the absence of any incident fluxes.
    However, these differences will not prevent us from obtaining a solution to Eq.~\eqref{eq:kin_eq_n2} by exploiting its similarity to Eq.~\eqref{eq:kin_eq_e}.

    To obtain a solution in the same manner as in  Sec.~\ref{sec:electrons}, 
    we require a small parameter that characterizes the interaction with the vortex.
    A naturally arising dimensionless ratio is%
    \footnote{
    The interpretation of the smallness of this parameter within the framework of the infinitely-thin vortex core model is discussed immediately following Eq.~\eqref{eq:Qpv_ineq}.
    }
    \begin{equation}
        \label{eq:Qpv_Pfn_rel}
        \frac{Q_{pv}}{p_{Fn}} \sim  \frac{p_{Fp}}{p_{Fn}} \epsilon,
    \end{equation}
    where the parameter $\epsilon$ is given by Eq.~\eqref{eq:eps_p_def}.
    Substituting Eq.~\eqref{eq:Nn_ans} into Eq.~\eqref{eq:kin_eq_n2} and retaining terms linear in the incident flux and up to quadratic order in the parameter $\epsilon$, we obtain
    \begin{align}
    	\vq \frac{\partial \tilde{g}_n }{ \partial \vr} 
	       -  \left[ \vq\times \Rot \left( m_n^* \gamma_{np} \vQ_{pv} \right) \right]  \frac{\partial \tilde{g}_n }{ \partial \vq}   
        =
	       -  \vVrn \left[ \vq\times \Rot \left( m_n^* \gamma_{np} \vQ_{pv} \right) \right]
	       - m_n^* \vVrn \frac{\partial \Delta \tilde{H}_{\vq,2}^{(n)} }{ \partial \vr},
	    \label{eq:dpn_eq2}
    \end{align}
    where Eqs.~\eqref{eq:en_ser2} and \eqref{eq:vgn_def} were accounted.
    Note that, since  the function $\tilde{g}_n$ is required only 
    to linear order in the incident fluxes, it is sufficient to retain only the part of $\Delta \tilde{H}_{\vq,2}^{(n)}$ that is independent of $\vVre$, $\vVrn$, and $\vQ_{p0}$.
    The general form of this part is
    \begin{equation}
    	\label{eq:DH2n_frm}
    	\Delta \tilde{H}_{\vq,2}^{(n)} = \mathcal{A} \vQ_{pv}^2 + \mathcal{B} (\vq \vQ_{pv})^2,
    \end{equation}
    where $\mathcal{A}$ and $\mathcal{B}$ are some functions of Landau parameters. 
    Thus, the first term on the right-hand side of Eq.~\eqref{eq:dpn_eq2} is of order $\epsilon$, whereas the second term is of order $\epsilon^2$.

    Eq.~\eqref{eq:dpn_eq2} is a linear first-order differential equation.
    Similar to Eq.~\eqref{eq:dpe_eq_tmp2}, it can be integrated along the characteristics.
    Therefore, its solution can generally be represented as a sum of two terms:
    \begin{equation}
        \label{eq:g_n_rep}
        \tilde{g}_{n} = \tilde{g}_{n,\rm scat.} + \tilde{g}_{n,H_2},
    \end{equation}
    where $\tilde{g}_{n,\rm scat.}$ and $\tilde{g}_{n,H_2}$ are contributions to the total solution, arising from the first and second terms on the right-hand side, respectively. 
    One can notice that Eq.~\eqref{eq:dpn_eq2}, with the second term on the right-hand side omitted, is equivalent to Eq.~\eqref{eq:dpe_eq_tmp2}, which describes the electron scattering. 
    Hence, $\tilde{g}_{n,\rm scat.}$ coincides with the function $g_e$ discussed in Sec.~\ref{sec:electrons}, with the magnetic field $\vB_v$ replaced by $(c/e) \Rot \left(m_n^*  \gamma_{np} \vQ_{pv}\right)$ and the momentum variable $\vp$ replaced by $\vq$.
    The explicit expression for the large-distance asymptotic of this term is given by Eq.~\eqref{eq:dpn_ld} below.
    To obtain the function $\tilde{g}_{n,H_2}$, we recall that the second term on the right-hand side of Eq.~\eqref{eq:dpn_eq2} is already of order $\epsilon^2$.
    Therefore, to the adopted level of accuracy, it is sufficient to integrate it along the unperturbed characteristics (i.e., along straight lines).
    Choosing again the boundary condition so that the function $\tilde{g}_{n,H_2}$ vanishes far upstream of the scattering region, we get
    \begin{equation}
        \label{eq:g_nH2_def}
       \tilde{g}_{n,H_2} =  -  \frac{m_n^*}{q_\perp}\,  \vVrn \int\limits_{-\infty}^{y_q} dy_1 \frac{\partial \Delta \tilde{H}_{\vq,2}^{(n)}(\vrd_1) }{ \partial \vrd_1},  
    \end{equation}
    where $\vq_\perp$ is the component of the vector $\vq$ perpendicular to $\vec{e}_z$.
    Here, we introduced the spatial coordinates defined relative to the vector $\vq$: 
    \begin{equation}
	   \label{eq:xy_def_ap}
		x_q = \rho \sin(\phi_q - \phi), \ \ \ \  y_q = \rho \cos(\phi_q - \phi),
	\end{equation}
    $\phi_q$ is the azimuthal angle of the vector $\vq$, and $\vrd_1 = (x_q, y_1)$.
    As will be demonstrated in what follows (see Sec.~\ref{sec:force} and Appendix~\ref{sec:ap:H2_contr}), 
    the function $\tilde{g}_{n,H_2}$ does not contribute to the force on the vortex. 
    Hence, it is not of interest to us in this paper.

    Returning to the first term of Eq~\eqref{eq:g_n_rep}, at distances $\rho \gg \lambda$, it has the form [cf.~Eq.~\eqref{eq:dpe_ld}]:
    \begin{equation}
    	\label{eq:dpn_ld}
        \tilde{g}_{n, \rm scat.} 
        =  \left[ - \vq \left(  \vec{e}_z \times \vVrn \right) \sigma_{\perp}^{(n)} +  \vq_\perp \vVrn  \sigma_{||}^{(n)} \right] \frac{\delta(\phi-\phi_\vq)}{\rho},
    \end{equation}
    where we introduced, respectively,  the transverse and transport cross sections for the neutrons: 
    \begin{equation}
        \label{eq:sigma_n_perp}
        \sigma_{\perp}^{(n)} =  \frac{m_p \varkappa}{ q_\perp} a_n,
    \end{equation}
    \begin{equation}
        \label{eq:sigma_n_par}
	   \sigma_{||}^{(n)} =  \frac{(m_p \varkappa)^2 }{q_\perp^2} L_n^{-1},
    \end{equation}
    and the following notations:
    \begin{equation}
    	\label{eq:an_def}
    	a_n = - m_n^{*} \gamma_{np}  \int\limits_{-\infty}^{\infty} dx \int \limits_{-\infty}^{\infty} dy\frac{\tilde{\mathcal{P}}'(\sqrt{x^2+y^2})}{2\pi \sqrt{x^2+y^2}},      
    \end{equation}
    \begin{equation}
    	\label{eq:Ln_def}
    	L_n^{-1} = (m_n^{*} \gamma_{np})^2 \frac{1}{2}\int\limits_{-\infty}^{\infty} dx \left( \int \limits_{-\infty}^{\infty} dy\frac{\tilde{\mathcal{P}}'(\sqrt{x^2+y^2})}{2\pi \sqrt{x^2+y^2}}   \right)^2.
    \end{equation}
    Eqs.~\eqref{eq:an_def} and \eqref{eq:Ln_def} are similar in form to Eqs.~\eqref{eq:ae_def} and \eqref{eq:Le_def}.
    Note, however, that the function $\mathcal{P}'(\rho)$, which determines the vortex magnetic field [see Eq.~\eqref{eq:B_vortex}], is replaced by the function - $m_n^{*} \gamma_{np} \tilde{\mathcal{P}}'(\rho)$, arising from $\Rot(m_n^{*} \gamma_{np} \vQ_p )$ [see Eq.~\eqref{eq:rotQpv}].

    Let us first consider the transverse cross section. 
    It is easy to see that the coefficient $a_n$ is proportional to the total flux $\Phi_Q$ [see Eq.~\eqref{eq:Phi_Q_def}; cf. Eq.~\eqref{eq:ae_val}].
    However, as it was argued in Sec.~\ref{sec:protons}, $\Phi_Q(\rho)$ tends to zero as $\rho \rightarrow \infty$.
    Therefore, the transverse cross section vanishes:
    \begin{equation}
        \label{eq:simga_n_perp_val}
        \sigma_{\perp}^{(n)} = 0.
    \end{equation}

    Inserting the function $\tilde{\mathcal{P}}'(\rho)$ given by Eq.~\eqref{eq:dPt_def} into 
    the function~\eqref{eq:Ln_def}, which controls the transport cross section \eqref{eq:sigma_n_par}, we find that it decomposes into three distinct terms:
    \begin{equation}
    	\label{eq:Ln_val}
    	L_n^{-1} = (m_n^{*} \gamma_{np})^2 L_e^{-1}
	       -   (m_n^{*} \gamma_{np})^2  \int \limits_{-\infty}^{\infty} dy\frac{{\mathcal{P}}'( {|y|})}{2\pi {|y|}}   
	       +    \frac{ (m_n^{*} \gamma_{np})^2}{2 } \int\limits_{-\infty}^{\infty} dx \left( \int \limits_{-\infty}^{\infty} dy\frac{\delta(\sqrt{x^2+y^2})}{2\pi \sqrt{x^2+y^2}}   \right)^2.
    \end{equation}
    Here, $L_e^{-1}$ is given by Eq.~\eqref{eq:Le_def}.
    In evaluating the second term of Eq.~\eqref{eq:Ln_val}, we made use of the identity~\eqref{eq:deltas_rel}.
    The third term formally diverges due to the presence of 
    two delta functions.
    This divergence indicates that the assumption of an infinitely thin vortex core is an oversimplification that prevents a complete calculation of the transport cross section.
    Therefore, to calculate this cross section, one must determine the structure of the vortex core by solving a self-consistent problem. 
    Nonetheless, it remains possible to make an estimate.

    A physical vortex has a core of  size $\sim \xi$, where the energy gap $\Delta_p$ drops to zero.
    On the other hand, the coefficient $\gamma_{np}$ vanishes when the superconductivity is destroyed \cite{GusakovHaensel2005}. 
    Therefore, somewhat loosely, the vortex core could be modeled 
    by introducing a spatially dependent coefficient
    $\gamma_{np} = \gamma_{np}^\infty \,  h(\rho)$,
    where $\gamma_{np}^\infty$ denotes its bulk value
    and $h(\rho)$ is some function smoothly
    rising from zero at the 
    vortex
    axis
    to unity 
    over a characteristic distance
    $\sim  \xi$.%
    \footnote{Note that in reality the dependence of the energy $\Delta H_{\vp}^{(n)}$ on $\vQ_p$ is nonlocal at the length scale of $\xi$.}
    In this case, we can still derive formally the same Eq.~\eqref{eq:kin_eq_n2} and, consequently, Eq.~\eqref{eq:dpn_eq2} but now $\Rot(m_n^* \gamma_{np} \vQ_p)$ is assumed to vanish at the vortex axis. 
    As a result, we again arrive at Eq.~\eqref{eq:dpn_ld} with the cross sections given by Eqs.~\eqref{eq:sigma_n_perp} and \eqref{eq:sigma_n_par},
    but with the corresponding coefficients now taking the form
    \begin{equation}
    	\label{eq:an_ns}
    	a_n = - m_n^{*} \gamma_{np}^\infty \int\limits_{-\infty}^{\infty} dx \int \limits_{-\infty}^{\infty} dy\frac{\mathcal{R}'(\sqrt{x^2+y^2})}{2\pi \sqrt{x^2+y^2}},      
    \end{equation}
    \begin{equation}
    	\label{eq:Ln_ns}
    	L_n^{-1} = \left( m_n^{*} \gamma_{np}^\infty \right)^2  \frac{1}{2}\int\limits_{-\infty}^{\infty} dx \left( \int \limits_{-\infty}^{\infty} dy\frac{\mathcal{R}'(\sqrt{x^2+y^2})}{2\pi \sqrt{x^2+y^2}}   \right)^2,
    \end{equation}
    where 
    \begin{equation}
        \label{eq:tPi_alt_def}
        {\mathcal{R}}' =     \frac{d}{d \rho} \left\{  h(\rho) \left[\mathcal{P}(\rho) - 1 \right] \right\}
                \approx \mathcal{P}'(\rho) - h'(\rho),
     \end{equation} 
     and $h'(\rho) = {dh}/{d \rho}$.
     The approximate equality in Eq.~\eqref{eq:tPi_alt_def} was obtained assuming that the function $h(\rho)$ varies much faster than the function $\mathcal{P}(\rho)$ (i.e.~$\xi \ll \lambda$).
     One can see that the approximate form of the function ${\mathcal{R}}'$ is similar to the function $\tilde{\mathcal{P}}'(\rho)$ given by Eq.\eqref{eq:dPt_def} but with the smooth function $h'(\rho)$ instead of the delta function.

    Being substituted into Eq.~\eqref{eq:an_ns}, the function \eqref{eq:tPi_alt_def} still gives $a_n = 0$.
    Considering the parameter $L_n^{-1}$ now given by Eq.~\eqref{eq:Ln_ns},
    one can verify that,
    under the assumption $\xi \ll \lambda$, the first two terms of the representation \eqref{eq:Ln_val} remain approximately the same except that the coefficient $\gamma_{np}$ should be replaced with $\gamma_{np}^\infty$.
    To estimate the third term, we adopt the following model function:
    \begin{equation}
        h(\rho) = 1- \exp\left( - \frac{\rho^2}{\xi^2}\right),
    \end{equation}
    which yields, instead of the third term in Eq.~\eqref{eq:Ln_val},   
    \begin{equation}
    	\label{eq:Ln_val2}
    	L_{n, \rm core}^{-1} =     \frac{\left( m_n^{*} \gamma_{np}^\infty \right)^2 }{2 } \int\limits_{-\infty}^{\infty} dx \left( \int \limits_{-\infty}^{\infty} dy\frac{h'(\sqrt{x^2+y^2})}{2\pi \sqrt{x^2+y^2}}   \right)^2 
        =
         \frac{1}{\sqrt{8\pi}}\frac{\left( m_n^{*} \gamma_{np}^\infty \right)^2 }{ \xi}.
    \end{equation}
    The numerical factor $1/\sqrt{8\pi}$ here depends on the specific choice of the model function $h(\rho)$
    and can change by a factor of few  depending on the choice of $h(\rho)$.
    However, the coefficient $L_{n, \rm core}^{-1}$, representing the core contribution to the total parameter $L_n^{-1}$,  
    is, in general, of order $(m_n^{*} \gamma_{np})^2/\xi$.
    On the other hand, taking into account that the function $\mathcal{P}'(\rho)$ vanishes at characteristic distance $\sim \lambda$, 
    the first two terms in Eq.~\eqref{eq:Ln_val} can be estimated as  $  (m_n^{*} \gamma_{np})^2/\lambda$.%
    \footnote{For the function $\mathcal{P}'(\rho)$ given by Eq.~\eqref{eq:dP_apr} these two integrals can be evaluated explicitly: the first one is given by Eq.~\eqref{eq:Le_val} and equals $1/8\lambda$, while the second one equals $1/2\lambda$. }
    Thus, we can conclude that, in the limit  $\xi \ll \lambda$, the third term produces the leading contribution to the neutron transport cross section.%

    Finally, let us return to the collision integral.
    Applying the same reasoning as in Sec.~\ref{sec:electrons} we derive the following condition [cf.~Eq.~\eqref{eq:le_cond}]:
    \begin{equation}
        \label{eq:ln_cond}
        \ell_n \gg  \frac{1}{ m_n^* \gamma_{np} }  \frac{p_{Fn} \lambda^2}{\hbar},
    \end{equation}
    which ensures the smallness of $I_n$.

\section{Forces on the vortex}
\label{sec:force}

    We now turn to calculating the forces using Eq.~\eqref{eq:F_mv_def}.
    Assuming the smallness of the incident particle fluxes, we restrict the calculation to linear order in $\vQ_{p0}$, $\vVre$, and $\vVrn$.
    It is most convenient to choose the radius of the integration cylinder $R$ such that $R \gg \lambda$.%
    %
    \footnote{Simultaneously, this radius should be much smaller than the intervortex distance.}
    %
    The contributions from the electrons, proton-neutron mixture, and electromagnetic field, can be calculated by substituting the corresponding part of the total stress tensor into Eq.~\eqref{eq:F_mv_def}.
    We begin with the electrons. 
    Inserting the distribution function \eqref{eq:Ne_ans} into the general expression \eqref{eq:Pi_e_ap} and extracting the contributions linear in the incident fluxes, we get
    \begin{equation}
    	\label{eq:Pi_e_lin}
    	\Pi_{e}^{ik} = \sum_{\vp\sigma}  p^i   \vvg^k ( g_e - e \varphi )\frac{\partial \mN_{\vp,0 }^{ (e)}}{\partial \varepsilon_\vp^{(e)}},
    \end{equation}	
    where $\mN_{\vp,0 }^{(e)} = f_{F}\left( \varepsilon_\vp^{(e)} - E_{Fe} \right)$ is the unperturbed electron distribution function.
    Here, we have taken into account that the electrostatic potential $\varphi$, generated by the scattering, is at least a linear function of $\vVre$, $\vVrn$, and $\vQ_{p0}$. 
    At the radius $R \gg \lambda$, we can make use of the approximate expression \eqref{eq:dpe_ld}.
    This allows us to write down the electron contribution to the force as
    \begin{equation}
	\label{eq:Fmv_e}
	\vec{F}_{\rm m \rightarrow v}^{(e)}
	= - \frac{1}{2} \sum_{\vp\sigma} v_{\vp\perp} p_\perp^2 \left[ -( \vec{e}_z \times\vVre) \sigma_\perp^{(e)} + \vVre \sigma_{||}^{(e)} \right] \frac{\partial \mN_{\vp,0 }^{ (e)}}{\partial \varepsilon_\vp^{(e)}}
	+ e  n_{e0} \int d^3 \vr \  \vec{E},
    \end{equation}
    where
    $v_{\vp\perp}  = p_\perp c^2 /\varepsilon_\vp^{(e)}$ and in the last term we made use of  Gauss’s theorem. 
    Approximating the derivative of the electron distribution function by the delta function:
    \begin{equation}
        \label{eq:dN_approx}
        \frac{\partial \mN_{\vp,0}^{(e)} }{\partial  \varepsilon_\vp^{(e)}} \approx - \delta \left(\varepsilon_\vp^{(e)} - E_{Fe} \right),
    \end{equation}
    and substituting Eqs.~\eqref{eq:sigma_e_perp}, \eqref{eq:sigma_e_par}, and \eqref{eq:ae_val},
    we obtain the electron contribution in the following form:
    \begin{equation}
    	\label{eq:Fmv_e2}
    	\vec{F}_{\rm m \rightarrow v}^{(e)}
        	= D'_e  \, \left( \vec{e}_z\times \vVre \right)
            + D_e \,  \vVre 
        	+ e \, n_{e0} \int d^3 \vr \  \vec{E},      
    \end{equation}
    where the following coefficients were introduced (cf.~Refs.~\cite{Gusakov2019,GG25}, see also Ref.~\cite{SourieChamel2020}):
    \begin{equation}
        D'_e = - m_p \varkappa \, n_{e0}
        \ \ \ \ 
        \mbox{and}
        \ \ \ \ 
        D_e = (m_p \varkappa)^2  \frac{3\pi}{8}   \frac{n_{e0}}{ p_{Fe}}  L_e^{-1},
    \end{equation}
    and $n_{e0} = {p_{Fe}^3}/{3\pi^2 \hbar^3}$.

    Let us now turn to the neutron-proton mixture. 
    The general expression for the neutron-proton stress tensor is given by Eq.~\eqref{eq:Pi_np_apv2}.
    It can be rewritten as follows:
    \begin{equation}
        \label{eq:Pi_np_v2}
        \Pi_{np}^{ik}  = 
    	   Q_p^i j_p^k
    	   - \sum_{\vp \sigma} \left( \varepsilon_\vp^{(n)}  - E_{Fn} \right) p^i  \frac{\partial \mN_{\vp}^{(n)} }{ \partial p^k}  
    	+ \delta^{ik} \left( n_p \breve{\mu}_p + n_n E_{Fn} - \mathcal E_{np} \right).
    \end{equation}
    Here,
    $\vj_p$ is the proton current given by Eq.~\eqref{eq:jp_def_ap}, and
    $\mathcal E_{np}$ is the energy of the neutron-proton subsystem given by Eq.~\eqref{eq:E_np}.
    To obtain this expression from Eq.~\eqref{eq:Pi_np_apv2}, we added and subtracted the combination $\sum_{\vp \sigma} E_{Fn}  p^i ( {\partial \mN_{\vp}^{(n)} }/{ \partial p^k} )$ and integrated one of these terms by parts.

    Our aim is to extract from Eq.~\eqref{eq:Pi_np_v2} the contribution linear in the incident fluxes far from the vortex. 
    At distances  $\rho \gg \lambda$, the vortex azimuthal flow  $\vQ_{pv}$ is exponentially suppressed.
    Therefore, the proton current can depend only on the incident flux and, consequently, the first term of Eq.~\eqref{eq:Pi_np_v2} is quadratically small.
    Next, we note that the neutron-proton energy depends on the incident fluxes through the neutron and proton distribution functions.
    Therefore, the linear contribution to the combination 
    multiplied by $\delta^{ik}$
    can be written as 
    \begin{equation}
        \label{eq:parent_var}
         \delta_{\rm l} \left( n_p \breve{\mu}_p + n_n E_{Fn} - \mathcal E_{np} \right) 
         \approx  
         - \sum_{\vp\sigma} \left\{ \left[ \left( \frac{\delta \mathcal E_{np}}{\delta \mN_{\vp}^{(n)}} \right)_{\rm eq,0} - E_{Fn}\right]  \delta_{\rm l} \mN_{\vp}^{(n)} 
            + \left[ \left(\frac{\delta \mathcal E_{np}}{\delta \mN_{\vp}^{(p)}}\right)_{\rm eq,0}  - E_{Fp} \right]  \delta_{\rm l} \mN_{\vp}^{(p)} \right\} + n_{p0}\,  \delta_{\rm l} \breve{\mu}_p.
    \end{equation}
    Here, the symbol $\delta_{\rm l}$ denotes the linear correction to a quantity, and the index ``${\rm eq},0$'' means that the corresponding variational derivative should be evaluated using equilibrium distribution functions with all currents set to zero ($\vVre=\vVrn=\vQ_p=0$).
    In deriving Eq.~\eqref{eq:parent_var}, we also took into account Eq.~\eqref{eq:mup_repr}.
    As follows from Eq.~\eqref{ddqw}, the proton contribution enclosed in the second square brackets vanishes.
    As for the neutrons, by definition [see Eq.~\eqref{eq:qp_energy_def}],
    \begin{equation}
        \left( \frac{\delta \mathcal E_{np}}{\delta \mN_{\vp}^{(n)}} \right)_{\rm eq,0} = \varepsilon_{\vp,0}^{(n)}.
    \end{equation}
    Therefore, the linear contribution to the stress tensor in Eq.~\eqref{eq:Pi_np_v2} takes the following form:
    \begin{equation}
        \label{eq:Pi_np_v2_lin}
         \delta_{\rm l}  \Pi_{np}^{ik}  = 
            - \sum_{\vp \sigma} \frac{ p^i p^k}{m_n^*} \frac{\partial  \mN_{\vp ,0}^{(n)}  }{\partial \varepsilon_{\vp,0}^{(n)}}  \delta_{\rm l} \varepsilon_\vp^{(n)}             
    	    - \sum_{\vp \sigma} \left( \varepsilon_{\vp,0}^{(n)}  - E_{Fn} \right) p^i  \frac{\partial }{ \partial p^k}    \delta_{\rm l} \mN_{\vp}^{(n)}
    	- \delta^{ik} \left(\sum_{\vp \sigma} \left( \varepsilon_{\vp,0}^{(n)}  - E_{Fn} \right)  \delta_{\rm l} \mN_{\vp}^{(n)}
        - n_{p0} \, \delta_{\rm l}  \breve{\mu}_p  \right),
    \end{equation}
    where $\mN_{\vp,0 }^{(n)} = f_{F}\left( \varepsilon_{\vp,0}^{(n)} - E_{Fn} \right)$.
    To proceed further, we notice that [see Eq.~\eqref{eq:Nn_ans_tmp}]
    \begin{equation}
        \delta_{\rm l} \mN_{\vp}^{(n)} =
       \frac{\partial \mN_{\vp,0}^{(n)}}{\partial  \varepsilon_{\vp,0}^{(n)}} 
        \left(\delta_{\rm l} \varepsilon_{\vp}^{(n)}  - \vp \vVrn + g_n  \right).
    \end{equation}
    Substituting this representation into Eq.~\eqref{eq:Pi_np_v2_lin}, we first observe that the terms $\propto \vp \vVrn$ vanish after averaging over the solid angle of the vector $\vp$.
    We then note that, at distances $\rho \gg \lambda$ and to linear order in the incident fluxes, one may simply replace the variable $\vp$ with $\vq$ [see Eq.~\eqref{eq:p_q_change}] and the function $g_n$ with $\tilde{g}_n$ [see Eq.~\eqref{eq:g_n_tilde_def}].
    After these manipulations, we arrive at
    \begin{equation}
    	\label{eq:Pi_np_fin}
    	\Pi_{np}^{ik}  = 
	    \sum_{\vq \sigma}    \tilde{g}_n   \frac{{q}^i  q^k}{m_n^*} \frac{\partial  \mN_{\vq ,0}^{(n)}  }{\partial \varepsilon_{\vq,0}^{(n)}} 
	       + \delta^{ik}   \, n_{p0}\,  \delta_{\rm l} \breve{\mu}_p.
    \end{equation}

    To calculate the proton-neutron mixture contribution to the force, one needs to substitute the function $\tilde{g}_n$ given by Eq.~\eqref{eq:g_n_rep}.
    It is shown in Appendix~\ref{sec:ap:H2_contr} that there is no contribution to the force from the function $\tilde{g}_{n,H_2}$.
    As for the function $\tilde{g}_{n,\rm scat.}$, its contribution has the form similar to Eq.~\eqref{eq:Fmv_e}.
    Thus, the total neutron-proton  contribution equals
    \begin{equation}
	\label{eq:Fmv_np}
	\vec{F}_{\rm m \rightarrow v}^{(np)}
	= - \frac{1}{2} \sum_{\vq\sigma} \frac{q_\perp^3}{m_n^*}  \left[ - (\vec{e}_z \times \vVrn) \sigma_\perp^{(n)} + \vVrn \sigma_{||}^{(n)} \right] \frac{\partial \mN_{\vq,0 }^{ (n)}}{\partial \varepsilon_{\vq,0}^{(n)}}
	- e n_{p0} \int d^3 \vr \  \vec{E}.
    \end{equation}
    Here, in the last term, we substituted Eq.~\eqref{eq:sf_eq} and applied the Gauss's theorem. 
    As in the case of electrons, we can approximate:
    \begin{equation}
        \label{eq:dNn_approx}
        \frac{\partial \mN_{\vq,0}^{(n)} }{\partial  \varepsilon_{\vq,0}^{(n)}} \approx - \delta \left(\varepsilon_{\vq,0}^{(n)} - E_{Fn} \right).
    \end{equation}   
    This allows us to obtain
    \begin{equation}
    	\label{eq:Fmv_np2}
    	\vec{F}_{\rm m \rightarrow v}^{(np)}
    	= 
         D_n  \vVrn      
    	- e n_{p0} \int d^3 \vr \  \vec{E},      
    \end{equation}
    where we introduced the coefficient
    \begin{equation}
        D_n = 
         (m_p \varkappa)^2 
        \frac{3\pi}{8}   \frac{n_{n0}}{ p_{Fn}}  L_n^{-1},
        \label{Dn11}
    \end{equation}
    and the unperturbed neutron density $n_{n0} = {p_{Fn}^3}/{3\pi^2 \hbar^3}$,
    and 
    substituted Eqs.~\eqref{eq:sigma_n_par} and \eqref{eq:simga_n_perp_val}.

    All that remains is to consider the contribution from the electromagnetic field. 
    The corresponding stress tensor is given by Eq.~\eqref{eq:Pi_EM_ap}.
    The electric field is linear in the incident fluxes; hence, the electric component of the tensor is quadratically small. 
    As for the magnetic contribution, apart from terms quadratic in the incident fluxes, it contains terms proportional to $ B_v^i B_v^k$ and $ B_v^i \delta B^k$. However, these terms are exponentially suppressed at large distances and can therefore be neglected. Consequently, the electromagnetic contribution vanishes within the accepted level of accuracy.

    Now we can write the total force acting on the vortex. 
    To this end, we notice that
    the terms proportional to electric field in Eqs.~\eqref{eq:Fmv_e2} and \eqref{eq:Fmv_np2} cancel each other due to the quasineutrality condition \eqref{eq:quasi}.
    Therefore, the total force equals 
    \begin{equation}
	\label{eq:Fmv}
	\vec{F}_{\rm m \rightarrow v}
	= D'_e  \, \left[ \vec{e}_z\times \left(\vVre - \vv\right) \right]
        - D_e \, \vec{e}_z \times \left[ \vec{e}_z \times \left(\vVre - \vv\right) \right]
        - D_n \, \vec{e}_z \times \left[ \vec{e}_z \times \left(\vVrn - \vv\right) \right].
    \end{equation}
    To make this expression more suitable for practical use, we switched to the laboratory frame of reference introducing the velocity of the vortex motion $\vv$.
    We also substituted $\vVra \rightarrow - \vec{e}_z \times \left( \vec{e}_z \times \vVra \right)$ 
    to allow for possible motion of the mixture constituents along the vortex lines.

    \section{Discussion and conclusions} 
    \label{sec:dis}

    Equation~\eqref{eq:Fmv}, with the coefficient $D_n$ given by Eq.\ \eqref{Dn11}, 
    constitutes the main result of this work.
    The total force given by this equation consists of three components. 
    The first two components are well known in the astrophysical literature. 
    They arise from the scattering of the electrons by the vortex magnetic field. 
    One can say that these components are
    electromagnetic in nature.%
    \footnote{Strictly speaking, this conclusion is fully valid only at length scales much larger than the London penetration depth. At smaller scales, the mechanism of momentum transfer becomes more intricate \cite{GG25}. }
    In contrast, 
    the third term cannot be reduced to an electromagnetic interaction.
    It describes neutron scattering and represents a manifestation of the
    nuclear forces.
    To investigate this force, we employed the Landau Fermi-liquid formalism.   
    This enabled us to treat the scattering of neutron quasiparticles in close analogy with the electron-scattering problem.
    Moreover, since the considered effect is a pure consequence of Fermi-liquid 
    interaction, 
    one expects the coefficient $D_n$ to be expressible in terms of the Landau parameters.
    Indeed, it is proportional to the square of the coefficient $\gamma_{np}$, introduced via Eq.~\eqref{eq:DH1_def}, which, in turn, in the case of strong proton superconductivity and normal neutrons, is given by \cite{GG23}%
    \footnote{In the cited reference the coefficient $\gamma_{np}$ is expressed through less common parameters $G_{\alpha\alpha'}$ which are connected with the standard Landau parameters via the equation 
    $ F_1^{\alpha\alpha'} / 3= \sqrt{ m_\alpha^* m_{\alpha'}^* / n_{\alpha0} n_{\alpha'0}}   G_{\alpha\alpha'}$.
    }
    \begin{equation}
        \gamma_{np} = \sqrt{\frac{1}{m_n^* m_p^*}} \sqrt{\frac{n_{p0}}{n_{n0}}} \frac{F_1^{np}/3}{ \left(1 + F_1^{nn}/3 \right)},
    \end{equation}
    where $F_1^{\alpha\alpha'}$ are the symmetric dimensionless Landau parameters defined for a Fermi-liquid mixture in the standard way \cite{PinesNozieres,GusakovHaensel2005,ChamelHaensel2006}.
    Therefore, this effect vanishes if the neutron-proton Fermi-liquid interaction is switched off ($F_1^{np}=0$).

    An important difference between the electron and neutron scattering is that the latter does not produce a transverse component of the force. The coefficient $D_e'$ governing the transverse electron force is proportional to the total magnetic flux generated by a proton vortex [see Eqs.~\eqref{eq:sigma_e_perp} and \eqref{eq:ae_val}].
    In the case of neutron scattering, the role of the magnetic field is played by the vector $\Rot(m_n^* \gamma_{np} \vQ_{pv})$.
    The total flux of this vector equals zero. 
    As argued in Sec.~\ref{sec:protons}, this is a consequence of the Meissner effect, which is universal  for superconductors  for any value of $\lambda/\xi$.
    Therefore, the absence of the neutron induced transverse force is expected to persist in the more realistic case of $\xi\sim \lambda$.

\begin{figure}
	\includegraphics[width=0.5\textwidth,trim= 1.0cm 0.0cm 1.0cm 1.0cm, clip = true]{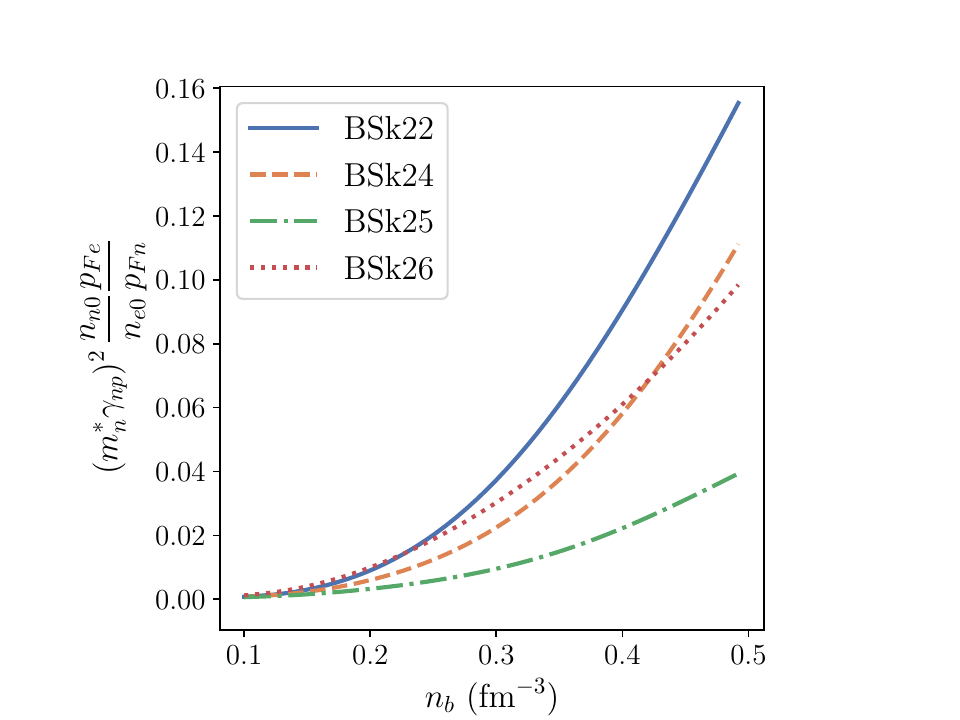}    
	\caption{
            \label{fig:coeff}
            The quantity $ (m_n^{*} \gamma_{np})^2 ({n_{n0}}/{n_{e0}}) ({p_{Fe}}/{p_{Fn}})$ plotted as a function of baryon number density $n_b$ for the BSk family of equations of state.
            The Landau parameters are calculated as described in Ref.~\cite{ChamelHaensel2006}.
          }
    \end{figure}

    Considering the longitudinal components,
    as we saw, the infinitely-thin vortex core model does not permit a complete calculation of the coefficient $D_n \sim L_n^{-1}$.
    To determine it, one must solve a self-consistent problem for the vortex-core structure.
    However, as was argued, it is natural to expect that $ L_n \sim (m_n^*\gamma_{np})^{-2} \xi$, and, consequently, one can make a physically motivated estimation:
    \begin{equation}
        \frac{D_n}{D_e} =  \frac{n_{n0}}{n_{e0}} \frac{p_{Fe}}{p_{Fn}} \frac{L_e}{L_n} 
        \approx \mathcal{K}\, (m_n^{*} \gamma_{np})^2 \frac{n_{n0}}{n_{e0}} \frac{p_{Fe}}{p_{Fn}} \frac{\lambda}{\xi}.
    \end{equation}
    where $\mathcal{K}$ is a dimensionless coefficient of the order of unity.
    The value of the ratio $n_{n0}/n_{e0}$ (or $p_{Fe}/p_{Fn}$) as well as the Landau parameters entering the coefficient $\gamma_{np}$ depend on the chosen equation of state. 
    Fig.~\ref{fig:coeff} demonstrates the value of the combination $ (m_n^{*} \gamma_{np})^2 ({n_{n0}}/{n_{e0}}) ({p_{Fe}}/{p_{Fn}})$ as a function of the baryon number density calculated for the BSk family of equations of state \cite{GorielyChamelPearson2013b,PearsonEtAl2019}.  
    As for the ratio $\lambda/\xi$, it is large in the regime of strong type-II superconductivity. 
    However, it remains bounded because the quasiclassical approximation adopted in the present work requires the quasiparticle wavelengths to be small compared with $\xi$. 
    Therefore, to maintain the consistency of the model, one must assume that $\lambda/\xi <  p_{Fn} \lambda / \hbar \approx$ 30 -- 50 \cite{GlampedakisAnderssonSamuelsson2011}.
    Finally, to estimate the relative magnitude of the third term in Eq.~\eqref{eq:Fmv}, one needs to compare the velocities $V_e$ and $V_n$ measured in the vortex frame of reference.
    This cannot be done in general.
    The relationship between neutron and electron velocities can depend substantially on the specific process under consideration.
    It should be emphasized that, irrespective of its magnitude, the new force may qualitatively alter the dynamics
    in the $npe$ mixture. 
    Indeed, if one sets $D_n = 0$ in Eq.~\eqref{eq:Fmv}, the equation $\vec{F}_{\rm m \rightarrow v} = 0$ has only a trivial solution $\vVre = \vv$.%
    \footnote{Strictly speaking, an electron velocity component parallel to the vortices is still allowed. However, a flow aligned with the vortices everywhere can hardly be realized in NS.}
    Therefore, a nonzero electron velocity can be generated only by ``external'' forces [see Eq.~\eqref{eq:force_bal}].
    In contrast, Eq.~\eqref{eq:Fmv} with a finite coefficient $D_n$ allows to maintain the difference $\vVre - \vVrn$ without any velocity independent ``external'' forces. 

    As was already mentioned in Sec.~\ref{sec:problem}, the strong type-II superconductivity approximation is  not particularly well suited for NS interiors. 
    Therefore, one can expect that the length scales $L_e$ and $L_n$ are of comparable magnitude. 
    However, this expectation should be verified through suitable calculations.
    Another limitation of treating the vortex core as infinitely thin is the inability to consistently account for the scattering of (quasi)particles by bound excitations \cite{KopninKravtsov1976b,SoninBook}.
    However, given the (quasi)particle large mean free path compared to  other length scales of the problem, one may expect the omitted processes to produce only a small correction.
    It is important to emphasize that the force discussed in the present work 
    cannot be reduced
    to the interaction of 
    neutrons
    with the 
    excitations bound in the vortex core.
    This is most clearly seen in the case of a strong type-II superconductor.
    Inspection 
    of Eq.~\eqref{eq:kin_eq_n2}
    shows that a neutron quasiparticle can sense the presence of the proton vortex at distances $\sim \lambda \gg \xi$, where no bound core excitations exist.
    
    In the present paper,
    we assumed that 
    the temperature is sufficiently low to completely neglect proton continuum excitations.
    Scattering of the proton quasiparticles is expected to contribute additional force components to Eq.~\eqref{eq:Fmv}. 
    The same applies if muons are present.   
    On the other hand, we assumed that the neutrons are normal (nonsuperfluid).
    If this condition is not satisfied, the neutron Bogoliubov excitations must be considered instead of the standard Landau quasiparticles. 
    Moreover, calculations indicate that the neutrons in the NS core form Cooper pairs in the triplet channel with an anisotropic energy gap \cite{SedrakianClark2019}.
    Although this would complicate the analysis, it is unlikely to qualitatively affect the results in the vicinity of the neutron critical temperature.
    If the temperature is far below $T_{cn}$, the number of neutron Bogoliubov excitations becomes exponentially 
    suppressed, and the force arising from neutron quasiparticle scattering correspondingly becomes negligible.
    It should also be noted that our analysis is restricted to the scattering of Fermi quasiparticles.
    Collective excitations may likewise contribute to the effect under consideration, particularly at low temperatures. 
    The primary aim of the present work is to demonstrate the existence of a force arising from the neutron scattering, which has previously been overlooked.
    Therefore, we consider the simplest set of conditions 
    that allow this effect to be isolated.
    Note, however, that the condition $T_{cn} \lesssim T \ll T_{cp}$ is not overly restrictive. 
    In the outer layers of NS cores, where type-II superconductivity is expected to occur \cite{GlampedakisAnderssonSamuelsson2011}, the proton critical temperature $T_{cp}$ is generally believed to exceed the neutron critical temperature $T_{cn}$ by 
    at least 
    a factor of 
    a few
    \cite{SedrakianClark2019}. 
    In turn, the regime $T_{cn}\lesssim T$ may be relevant for several classes of neutron stars of observational interest. 
    For example, the inferred internal temperatures of magnetars 
    (e.g., Refs.~\cite{kaminker_etal06,kkpy14,bl15,moraga_etal25}), 
    young isolated cooling neutron stars 
    (e.g., Refs.~\cite{fortin_etal18,wbs20,potekhin_etal20}), 
    and accreting neutron stars in LMXBs 
    (e.g., Refs.~\cite{ho11,gck14,brown_etal18,pcc19}) 
    can be comparable to, or even exceed, the density-dependent critical temperature for triplet neutron pairing 
    in some regions 
    of the core.


    In conclusion, we have 
    argued that the neutron-proton Fermi-liquid interaction 
    gives rise to a force proportional to the neutron velocity relative to the proton vortices.
    In this work, we employ a relatively simple model.
    Nevertheless, the main properties of the force, as well as physically motivated estimates, can already be extracted from the
    analysis presented here.
    The impact of this force on each specific process (magnetic field evolution, stellar oscillations, etc.) warrants separate investigation.

    \section*{Acknowledgments}
    This research was supported by the Russian Science Foundation Grant No. 22-12-00048-P.

\appendix

\section{Energy} 

\label{sec:energy}

    The total energy density of the system can be represented as 
    \begin{equation}
    	\label{eq:E_tot}
    	\mathcal E = \mathcal E_{np} + \mathcal E_e + \mathcal E_{EM},
    \end{equation}
    where  $\mathcal E_{np}$ is the energy of the neutron-proton subsystem, $ \mathcal E_e$ is the energy of the electrons, and $\mathcal E_{EM}$ is the energy of the electromagnetic field.
    The energy of the neutron-proton subsystem, in turn, is assumed to be a sum of two terms \cite{GusakovHaensel2005,GG23}:
    \begin{equation}
    	\label{eq:E_np}
    	\mathcal E_{np}  = \mathcal E_{\rm LF} + \mathcal E_{\rm pair}.
    \end{equation}
    In this expression, the first term formally coincides with the energy of a normal (nonsuperfluid) Fermi-liquid mixture \cite{PinesNozieres}:
    \begin{equation}
	    \label{eq:E_LF}
		\mathcal E_{\rm LF} =
			E_0   
			+ \sum_{\vp \sigma \alpha} \varepsilon_0^{(\alpha)}(\vp)    \,
			\left( \mN_{\vp}^{(\alpha)} - \theta_\vp^{(\alpha)} \right) 
			+ \frac{1}{2}  \sum_{\vp \vps  \sigma \sigma ' \alpha \alpha '}
			f^{\alpha \alpha'}\left(\vp , \vps  \right) \left( \mN_{\vp }^{(\alpha)} - \theta_\vp^{(\alpha)} \right)
			\left( \mN_{\vps  }^{(\alpha')} - \theta_\vps^{(\alpha')} \right)	,
    \end{equation} 
    where 
    $\mN_{\vp}^{(\alpha)}$ are the distribution functions of Landau quasiparticles  for species $\alpha = n, p$, 
    $\theta_\vp^{(\alpha)} = \theta(p_{F\alpha} - p)$,
    $\theta(x)$ is the Heaviside step function,
    $\varepsilon_{0}^{(\alpha)} \left(\vp \right)$ and $f^{\alpha \alpha'} \left(\vp, \vps \right)$ are the first and second variations of the energy 
    $\mathcal E_{\rm LF}$ 
    with respect to the distribution function $\mN_{\vp}^{(\alpha)}$ evaluated at the ground (nonsuperfluid) state $\mN_{\vp}^{(\alpha)} = \theta_\vp^{(\alpha)}$.
    We consider spin-saturated matter, and therefore omit spin indices from the distribution functions $\mN_{\vp}^{(\alpha)}$ while by $f^{\alpha \alpha'} \left(\vp, \vps \right)$ we mean the spin-averaged interaction function. 
    However, we retain formal summations over the spin  indices $\sigma, \sigma'$, etc., which essentially means multiplication by 2.

    The pairing energy term  in Eq.~\eqref{eq:E_np} equals \cite{deGennes_book}
    \begin{equation}
    	\label{eq:E_pair}
    	\mathcal E_{\rm pair} =   \frac{\Delta^{(p)2}}{\nu},
    \end{equation}
    where $\Delta^{(p)}$ denotes the proton energy gap
    and $\nu<0$ is the effective coupling constant.
    We assume  isotropic pairing for the protons. 
    Following the BCS approximation we ignore the dependence of the coupling constant and, consequently, of the gap on the absolute value of the momentum variable.

    Eq.~\eqref{eq:E_LF} allows us to define the Landau quasiparticle energy:
    \begin{equation}
    	\label{eq:qp_energy_def}
    	\varepsilon_\vp^{(\alpha)} = \frac{\delta \mathcal E_{\rm LF}}{\delta \mN_{\vp}^{(\alpha)} } 
    	=  \varepsilon_0^{(\alpha)}(\vp) +  \sum_{\vps   \sigma'   \alpha' } f^{\alpha \alpha'}\left(\vp , \vps  \right) \left( \mN_{\vp' }^{(\alpha')} - \theta_\vps^{(\alpha')}  \right).
    \end{equation}
    For the superconducting protons the following quantities can also be introduced:
    \begin{equation}
    	\label{eq:Hq_def}
    	\mathcal{H}^{(p)}_\vq = \frac{\varepsilon_{\vq + \vQ_p}^{(p)} + \varepsilon_{-\vq + \vQ_p}^{(p)} }{2} - \breve{\mu}_p,
    \end{equation}	
    \begin{equation}
    	\label{eq:Eq_def}
    	{E}^{(p)}_\vq = \sqrt{\mathcal{H}^{(p)2}_\vq+ \Delta^{(p)2}}.
    \end{equation}
    Note that, in the presence of the proton supercurrent $\sim \vQ_p$,
    it is more convenient to describe protons in terms of the momentum variable $\vq$ defined via Eq.~\eqref{eq:Hq_def}.
    This variable
    should
    not be confused with the shifted momentum variable $\vq$ defined via Eq.~\eqref{eq:p_q_change} for neutrons. 
    The function ${E}^{(p)}_\vq$ represents the symmetric part (${E}^{(p)}_\vq = {E}^{(p)}_{-\vq}$) of the proton Bogoliubov excitation energy [cf.~Eq.~\eqref{eq:bog_ex}].

    The following self-consistency relation should be satisfied for the energy gap:
    \begin{equation}
    	\label{eq:Delta_p_def}
    	1 = - \frac{\nu}{4}  \sum_{\vq,\sigma}  \frac{1}{{E}^{(p)}_\vq }.
    \end{equation}
    On the other hand, the energy gap can be expressed as follows: 
    %
    \begin{equation}
        \label{eq:delta_rep}
        \Delta^{(p)} = 2 {E}^{(p)}_\vq \uuq \vvq,
    \end{equation}
    where
    \begin{equation}
        \label{eq:UqVq_def}
        \uuq = \sqrt{\frac{1}{2}\left(1 + \frac{\mathcal{H}^{(p)}_\vq}{{E}^{(p)}_\vq} \right)}, \ \ \ \ \ 
        \vvq = \sqrt{\frac{1}{2}\left(1 - \frac{\mathcal{H}^{(p)}_\vq}{{E}^{(p)}_\vq} \right)}
    \end{equation}
    are the Bogoliubov coherence factors 
    related by the normalization condition 
    \begin{equation}
        \label{eq:uv_rel}
        \uuqsq  + \vvqsq = 1.
    \end{equation}
    In the limit of vanishing temperature, the proton distribution function 
    can be expressed through the following equation \cite{deGennes_book,GusakovHaensel2005}:
    \begin{equation}
        \label{eq:N_q_p}
        \mN_{\vq+\vQ_p}^{(p)} = \mN_{- \vq+\vQ_p}^{(p)} = \vvqsq.
    \end{equation}
   It is easy to check (see Appendix~\ref{sec:pvar}) that, in this limit, the following variational derivative vanishes,
    \begin{equation}
        \frac{\delta( \mathcal{E}_{np}-\breve{\mu}_p n_p)}{\delta \mN_{\vq+\vQ_p}^{(p)}}=\frac{\delta( \mathcal{E}_{np}-\breve{\mu}_p n_p)}{\delta \vvqsq}= 0,
        \label{ddqw}
    \end{equation}
   thereby ensuring that $\vvq$
   is related to the energy gap by Eq.~\eqref{eq:UqVq_def}.
  Note that, when varying \eqref{ddqw}, the chemical potential $\breve{\mu}_p$ must be treated as constant.

    The energy  of the electrons can be represented as
    \begin{equation}
        \mathcal E_{e} = \sum_{\vp\sigma}\varepsilon_\vp^{(e)}  \mN_{\vp}^{ (e)}.
    \end{equation}
    The energy of the electromagnetic field is given by the standard expression \cite{LL2}:
    \begin{equation}
    	\mathcal E_{EM} = \frac{1}{8\pi} (E^2 + B^2).
    \end{equation}

\section{Proof of Eq.\ \eqref{ddqw}}
\label{sec:pvar}

    In this appendix, we prove Eq.~\eqref{ddqw}.
    The variation $\delta(\mathcal E_{np}-\breve{\mu}_p n_p)$ with respect to the proton distribution function $\mN_{\vp}^{(p)}$, taken under the assumption that $\breve{\mu}_p$ is fixed, can be decomposed into two parts [see Eq.~\eqref{eq:E_np}]:
    \begin{align}
        \delta(\mathcal E_{np} - \breve{\mu}_p n_p)=\delta(\mathcal E_{\rm LF} - \breve{\mu}_p n_p)+\delta \mathcal{E}_{\rm pair}.
         \label{varw}
    \end{align}
  Taking into account 
  Eq.~\eqref{eq:qp_energy_def} and substituting $\vp=\vq+\vQ_p$, the first term in Eq.~\eqref{varw} can be represented as
    \begin{equation}
        \label{eq:Epn_var}
            \delta \left(\mathcal E_{\rm LF} - \breve{\mu}_p n_p \right)
            = \sum_{\vq\sigma} \left(\varepsilon_{\vq + \vQ_p}^{(p)} - \breve{\mu}_p \right) \delta \mN_{\vq + \vQ_p}^{(p)}.
    \end{equation}
    We restrict our consideration to functions $\delta \mN_{\vq+\vQ_p}^{(p)}$ that are invariant under the transformation $\vq \rightarrow -\vq$. This reflects the fact that Cooper pairing occurs between particles with momenta $\vq+\vQ_p$ and $-\vq+\vQ_p$.
    Accounting for this symmetry, Eq.~\eqref{eq:Epn_var} can be rewritten as
    \begin{equation}
        \label{eq:Epn_var2}
        \delta \left(
        \mathcal E_{\rm LF}
        - \breve{\mu}_p n_p \right)
            = \sum_{\vq\sigma} \mathcal{H}^{(p)}_\vq \delta \mN_{\vq + \vQ_p}^{(p)},
    \end{equation}
    where the function $\mathcal{H}^{(p)}_\vq$ is defined according to Eq.~\eqref{eq:Hq_def}.

    Let us now turn to the pairing energy term in Eq.\ \eqref{varw}.
    In view of Eqs.~\eqref{eq:Delta_p_def}, \eqref{eq:delta_rep}, and \eqref{eq:uv_rel}, we have
\begin{equation}
    \delta \mathcal E_{\rm pair}
    = - \frac{1}{2} \sum_{\vq\sigma} \Delta^{(p)} \, \delta \left( \frac{\Delta^{(p)}}{E^{(p)}_\vq} \right)
    = - \sum_{\vq\sigma} \Delta^{(p)} \, \frac{\uuqsq - \vvqsq}{\uuq } \, \delta \vvq,
\end{equation}
where in the first equality we used the fact that the variation of the right-hand side of Eq.~\eqref{eq:Delta_p_def} vanishes.
    The function $\vvq$ is related to the proton distribution function via Eq.~\eqref{eq:N_q_p}.
    Taking this into account and substituting Eqs.~\eqref{eq:UqVq_def}, we obtain
    \begin{equation}
        \delta \mathcal E_{\rm pair}  
           = -  \sum_{\vq\sigma}   \mathcal{H}^{(p)}_\vq \delta \mN_{\vq + \vQ_p}^{(p)}.
    \end{equation}
    Thus, the total variation in Eq.~\eqref{varw} vanishes.
Note that, in deriving this result, we did not assume that the whole system is in thermodynamic equilibrium with all currents set to zero.
If the variation is evaluated in such an equilibrium, one should replace
$\varepsilon_{\vq+\vQ_p}^{(p)} \rightarrow \varepsilon_{\vq,0}^{(p)}$
and $\breve{\mu}_p \rightarrow E_{Fp}$.

\section{Momentum conservation}
\label{sec:mom_cons}

    To derive the momentum conservation equation, let us write down the 
    time-dependent
    kinetic equations for neutrons and electrons:
    \begin{equation}
	\label{eq:kin_eq_n_ap}
	\frac{\partial \mN_{\vp}^{ (n)}}{\partial t} +
	\frac{\partial \varepsilon_\vp^{(n)}}{ \partial \vp} \frac{\partial \mN_{\vp}^{(n)} }{ \partial \vr} 
	- \frac{\partial \varepsilon_\vp^{(n)} }{ \partial \vr} \frac{\partial \mN_{\vp}^{(n)} }{ \partial \vp}	 = I_n,
    \end{equation}	
    \begin{equation}
	\label{eq:kin_eq_e_ap}
				\frac{\partial \mN_{\vp}^{ (e)}}{\partial t} + 
				\vvg \frac{\partial \mN_{\vp}^{ (e)}}{\partial \vr} 
				- e \left( \vec{E} +\frac{1}{c} \vvg \times \vec{B} \right) \frac{\partial \mN_\vp^{ (e)}}{\partial \vp} = I_e. 
    \end{equation}
    The proton condensate satisfies the ``superfluid" equation \cite{AronovEtAl1981,Gusakov2010}  
    \begin{equation}
	\label{eq:superfluid_eq_ap}
	\frac{\partial \vQ_p}{\partial t} + \nabla  \breve{\mu}_p  =   e   \vec E
    \end{equation}
    and the continuity equation
    \begin{equation}
	\label{eq:cont_eq_p_ap}
	\frac{\partial n_p}{\partial t} + \Div  \vj_p = 0.
    \end{equation}
    Here, the proton particle number density
    equals \cite{deGennes_book,GusakovHaensel2005}
    \begin{equation}
    	\label{eq:np_def_ap}
	       n_p = \sum_{\vq \sigma}   \mN_{\vq + \vQ_p}^{ (p)},
    \end{equation}
    where 
    the  distribution function is given by Eq.~\eqref{eq:N_q_p},
    while the proton current density can be expressed as \cite{Leggett1965,AllardChamel2021}
    \begin{equation}
	\label{eq:jp_def_ap}
	\vj_p = \sum_{\vp \sigma} \frac{\partial \varepsilon_\vp^{(p)}}{ \partial \vp} \mN_{\vp }^{ (p)} = \sum_{\vq \sigma} \frac{\partial \varepsilon_{\vq + \vQ_p}^{(p)}}{ \partial \vq} \mN_{\vq + \vQ_p }^{ (p)}.
    \end{equation}

    Let us multiply Eq.~\eqref{eq:kin_eq_n_ap} by $\vp$ and sum the result over all momenta $\vp$ and spin states $\sigma$. 
    This yields, after few integrations by parts:
    \begin{equation}
	\label{eq:mom_eq_tmp1}
	\frac{\partial }{\partial t} \sum_{\vp \sigma} p^i \mN_{\vp}^{ (n)} + \frac{\partial  }{ \partial r^k}  \sum_{\vp \sigma} p^i \frac{\partial \varepsilon_\vp^{(n)}}{ \partial p^k} \mN_{\vp}^{(n)}
	+  \sum_{\vp \sigma} \frac{\partial  \varepsilon_\vp^{(n)} }{ \partial r^i} \mN_{\vp}^{ (n)}
	=  \sum_{\vp \sigma} p^i I_n.
    \end{equation}
    The third term on the left-hand side does not generally vanish due to the Fermi-liquid interaction.
    To proceed further, let us calculate the gradient of the energy $\mathcal E_{LF}$.
    Using its definition 
    given by
    Eq.~\eqref{eq:E_LF},
    one can show that
    \begin{equation}
	 \nabla \mathcal E_{\rm LF}
	 = \sum_{\vp \sigma \alpha = n,p}   \left[   \nabla \left(   \varepsilon_\vp^{(\alpha)}  \mN_{\vp}^{ (\alpha)} \right)  -   \mN_{\vp}^{ (\alpha)}  \nabla \varepsilon_\vp^{(\alpha)}\right].
    \end{equation}
    Therefore, Eq.~\eqref{eq:mom_eq_tmp1} transforms to
    \begin{equation}
	    \label{eq:mom_eq_tmp2}
    	\frac{\partial }{\partial t} \sum_{\vp \sigma} p^i \mN_{\vp}^{ (n)} 
    	+ \frac{\partial  }{ \partial r^k} \left[ \sum_{\vp \sigma}  p^i \frac{\partial \varepsilon_\vp^{(n)}}{ \partial p^k} \mN_{\vp}^{(n)} 
    	+ \delta^{ik} \left(\sum_{\vp \sigma\alpha = n,p}  \varepsilon_\vp^{(\alpha)}  \mN_{\vp}^{ (\alpha)} - \mathcal E_{\rm LF} \right)
    	\right]
	       -  \sum_{\vp \sigma} \frac{\partial  \varepsilon_\vp^{(p)} }{ \partial r^i} \mN_{\vp}^{ (p)}
	       =  \sum_{\vp \sigma} p^i I_n.
    \end{equation}
    The last term on the left-hand side can be rewritten as
    \begin{equation}
    	\label{eq:mom_eq_tmp3}
    	\sum_{\vp \sigma}   \mN_{\vp}^{ (p)}  \nabla   \varepsilon_\vp^{(p)} 
    	= \sum_{\vq \sigma}   \mN_{\vq + \vQ_p}^{ (p)} \nabla \left(  \varepsilon_{\vq + \vQ_p}^{(p)}  - \breve{\mu}_p \right) - j_p^k \nabla Q_p^k +         n_p \nabla \breve{\mu}_p.
    \end{equation}
    To obtain this identity, we shifted the running momentum variable via the substitution $\vp = \vq + \vQ_p$ and then applied Eqs.~\eqref{eq:np_def_ap} and \eqref{eq:jp_def_ap}.
    To proceed further, one can show that  (cf.~Appendix \ref{sec:pvar})
    \begin{equation}
        \label{eq:NE_rel_tmp}
         \sum_{\vq \sigma}   \left(  \varepsilon_{\vq + \vQ_p}^{(p)}  - \breve{\mu}_p \right) \nabla \mN_{\vq + \vQ_p}^{ (p)} 
          = 
          \sum_{\vq \sigma} \mathcal{H}^{(p)}_\vq \nabla \vvqsq
          =
          - \nabla  \mathcal E_{\rm pair}.
    \end{equation}
    Using Eq.~\eqref{eq:NE_rel_tmp}, one can obtain the following expression for the first term on the right-hand side of Eq.\ \eqref{eq:mom_eq_tmp3}: 
    \begin{equation}
    	\sum_{\vq \sigma}   \mN_{\vq + \vQ_p}^{ (p)} \nabla \left(  \varepsilon_{\vq + \vQ_p}^{(p)}  - \breve{\mu}_p \right)
    	= \nabla \left[  \sum_{\vq \sigma}  \left(  \varepsilon_{\vq + \vQ_p}^{(p)}  - \breve{\mu}_p \right) \mN_{\vq + \vQ_p}^{ (p)} +  \mathcal E_{\rm pair} \right]
    	= \nabla \left[  \sum_{\vp \sigma}   \varepsilon_{\vp}^{(p)} \mN_{\vp}^{ (p)}  - n_p \breve{\mu}_p   +  \mathcal E_{\rm pair} \right].
       \label{333}
    \end{equation}
    The gradient of the chemical potential in the third term of Eq.~\eqref{eq:mom_eq_tmp3} can be expressed with the help of Eq.~\eqref{eq:superfluid_eq_ap}.
    Substituting the resulting expression for the term $\sum_{\vp \sigma}   \mN_{\vp}^{ (p)}  \nabla   \varepsilon_\vp^{(p)}$ back into Eq.~\eqref{eq:mom_eq_tmp2}, and making use of the continuity equation \eqref{eq:cont_eq_p_ap}, we get
    \begin{align}
	\nonumber
        \frac{\partial  }{\partial t}  \left( \sum_{\vp \sigma} p^i \mN_{\vp}^{ (n)} + n_p Q_p^i \right)
	+ Q_p^i \frac{\partial j_p^k}{\partial r^k} +  j_p^k \frac{\partial  Q_p^k}{\partial r^i}
	+ \frac{\partial  }{ \partial r^k} \left[ \sum_{\vp \sigma}  p^i \frac{\partial \varepsilon_\vp^{(n)}}{ \partial p^k} \mN_{\vp}^{(n)} 
	+ \delta^{ik} \left(\sum_{\vp \sigma}  \varepsilon_\vp^{(n)}  \mN_{\vp}^{ (n)} + n_p \breve{\mu}_p - \mathcal E_{np} \right)
	\right]
	\\
	- e n_p {E}^i
	=  \sum_{\vp \sigma} p^i I_n.
	\label{eq:mom_eq_tmp4}
    \end{align}
    Next, we can add and subtract the combination  $  j_p^k \frac{\partial Q_p^i }{\partial r^k}$  and then substitute Eq.~\eqref{eq:rotQ_ch}. 
    This leads us to the following equation:
    \begin{align}
	\frac{\partial \, \mathfrak{P}_{np}^i }{\partial t} 
	+ \frac{\partial \Pi_{np}^{ik}  }{ \partial r^k} 
	= e  \left( n_p {E}^i +\frac{1}{c} [\vj_p \times \vB]^i \right)
	+ \delta^{(2)}(\vrd)\, {F}_{\rm ext}^i
	 +  \sum_{\vp \sigma} p^i I_n,
	\label{eq:mom_eq_tmp5}
    \end{align}
    where we defined the neutron-proton momentum density
    \begin{equation}
        \mathfrak{P}_{np}^i =  \sum_{\vp \sigma} p^i \mN_{\vp}^{ (n)} + n_p Q_p^i
    \end{equation}
    and the neutron-proton stress tensor
    \begin{equation}
        \label{eq:Pi_np_ap}
        \Pi_{np}^{ik}  = 
    	Q_p^i j_p^k
    	+ \sum_{\vp \sigma}  p^i \frac{\partial \varepsilon_\vp^{(n)}}{ \partial p^k} \mN_{\vp}^{(n)} 
    	+ \delta^{ik} \left(\sum_{\vp \sigma}  \varepsilon_\vp^{(n)}  \mN_{\vp}^{ (n)} + n_p \breve{\mu}_p - \mathcal E_{np} \right).
    \end{equation}
    Performing some integrations by parts,
    the stress tensor can be rewritten in a form that is more in line with the Fermi-liquid framework:
    \begin{equation}
        \label{eq:Pi_np_apv2}
        \Pi_{np}^{ik}  = 
    	   Q_p^i j_p^k
    	   - \sum_{\vp \sigma}  \varepsilon_\vp^{(n)}    p^i  \frac{\partial \mN_{\vp}^{(n)} }{ \partial p^k}  
    	+ \delta^{ik} \left( n_p \breve{\mu}_p  - \mathcal E_{np} \right).
    \end{equation}
    The first term on the right-hand side of Eq.~\eqref{eq:mom_eq_tmp5} is the standard Lorentz force.
    In the second term, we introduced the quantity $\vec{F}_{\rm ext}$ via the equation
    \begin{equation}
        \label{eq:F_ext_def}
        \delta^{(2)}(\vrd)\, 
        \vec{F}_{\rm ext}
        = \frac{m_p \varkappa}{2 \pi \rho}\, \delta(\rho) \, \vec{e}_z\times \vj_p,
    \end{equation}
    in which
    the identity \eqref{eq:deltas_rel} was taken into account. 
    This delta-function containing term arises from taking the $\Rot$ of the condensate momentum $\vQ_p$ [see Eq.~\eqref{eq:rotQ_ch}].
    As shown in Ref.~\cite{GG25}, the quantity $\vec{F}_{\rm ext}$ defined in Eq.~\eqref{eq:F_ext_def} can be interpreted as the external force acting on an infinitely thin vortex core.
    We do not discuss this force further here and refer the reader to Ref.~\cite{GG25} for a detailed consideration.
    The third term on the right-hand side of Eq.~\eqref{eq:mom_eq_tmp5} describes the exchange of momentum between neutrons and electrons via collisions.

    Let us now turn to the electrons. 
    Multiplying Eq.~\eqref{eq:kin_eq_e_ap} by $\vp$ and summing it over $\vp$ and $\sigma$,
    we obtain
    \begin{align}
    	\frac{\partial \, \mathfrak{P}_{e}^i }{\partial t} 
    	+ \frac{\partial \Pi_{e}^{ik}  }{ \partial r^k} 
    	= - e  \left(n_e \vec{E} +\frac{1}{c}\vj_e \times \vec{B}\right)
	      +  \sum_{\vp \sigma} p^i I_e,
	\label{eq:mom_eq_e_tmp}
    \end{align}
    where 
    \begin{equation}
        \mathfrak{P}_{e}^i =  \sum_{\vp \sigma} p^i \mN_{\vp}^{ (e)}
    \end{equation}
    is the electron momentum density,
    \begin{equation}
        \label{eq:Pi_e_ap}
    	\Pi_{e}^{ik} = \sum_{\vp\sigma}  p^i   \vvg^k  \mN_{\vp}^{ (e)}
    \end{equation}
    is the electron stress tensor and $\vec{j}_e = \sum_{\vp\sigma} \vvg \mN_{\vp}^{ (e)} $ is the electron current density.
    Finally, we sum Eqs.~\eqref{eq:mom_eq_tmp5} and \eqref{eq:mom_eq_e_tmp}, and rewrite the total Lorentz force in terms of the derivatives of the electromagnetic momentum density
    \begin{equation}
        \mathfrak{P}_{\rm EM}^i = \frac{(\vec{E}\times\vec{B})^i}{4\pi c}
    \end{equation}
    and the electromagnetic stress tensor 
    \begin{equation}
    	\label{eq:Pi_EM_ap}
    	\Pi_{\rm EM}^{ik} = - \frac{1}{4\pi} \left[ E^iE^k + B^i B^k - \frac{\delta^{ik}}{2}\left( E^2 + B^2\right) \right]
    \end{equation}
    in the standard manner \cite{LL2}. 
    This leads us to Eq.~\eqref{eq:mom_eq_ap} with the total momentum density 
    \begin{equation}
        \label{eq:total_mom}
        \mathfrak{P}^i = \mathfrak{P}_{np}^i + \mathfrak{P}_{e}^i + \mathfrak{P}_{\rm EM}^i
    \end{equation}
    and total stress tensor 
    \begin{equation}
        \label{eq:total_stress}
         \Pi^{ik}  = \Pi_{np}^{ik}  +  \Pi_{e}^{ik} +  \Pi_{\rm EM}^{ik}.
    \end{equation}

\section{Contribution to the force from the function $\tilde{g}_{n,H_2}$}
\label{sec:ap:H2_contr}

    Substituting Eq.~\eqref{eq:g_nH2_def} into the stress tensor~\eqref{eq:Pi_np_fin} and, then, the stress tensor into Eq.~\eqref{eq:F_mv_def}, we can represent the result in the following form:
    \begin{equation}
        \label{eq:ap:F_H2}
        {F}_{\rm m \rightarrow v}^{(np)i} = 
            \frac{2}{(2\pi \hbar)^3} \int\limits_0^\infty dq\, q^2 
             \frac{\partial \mN_{\vq,0 }^{ (n)}}{\partial \varepsilon_\vq^{(n)}} 
             \int\limits_{4\pi} d\Omega_q\, q^i
             \int\limits_0^{2\pi} d\phi \,
             y_q
             \left[ \frac{\vVrn \vq_\perp}{q_\perp} \Delta \tilde{H}_{\vq,2}^{(n)}(x_q,y_q) -  \vVrn \frac{\vec{e}_z \times \vq_\perp}{q_\perp}  \int\limits_{-\infty}^{y_q} dy_1 \frac{\partial  \Delta \tilde{H}_{\vq,2}^{(n)}(x_q,y_1)}{\partial x_q} 
             \right].
    \end{equation}
    In this equation, 
    we replaced the summation over quantum states $(\vq,\sigma)$ by the integration over $\vq$ variable multiplied by 2, $d\Omega_q$ denotes a solid-angle element of the vector $\vq$.
    We also employed the coordinates $x_q$ and $y_q$ defined in Eq.~\eqref{eq:xy_def_ap} with $\rho$ set equal to $R$, the radius of the cylindrical integration surface.
    The quantity $\Delta \tilde{H}_{\vq,2}^{(n)}$ given by Eq.~\eqref{eq:DH2n_frm} can be represented as 
    %
    \begin{equation}
    	\label{eq:DH2n_ap}
    	\Delta \tilde{H}_{\vq,2}^{(n)}(x_q,y_q) = \left( \mathcal{A}  + q_\perp^2 \mathcal{B} \frac{x_q^2}{{x_q^2 + y_q^2}} \right)  Q_{pv}^2   \left( \sqrt{x_q^2 + y_q^2} \right).
    \end{equation}
    Let us consider the integral 
    \begin{equation}
        \mathcal{I} =  \int\limits_0^{2\pi} d\phi \,  y_q \int\limits_{-\infty}^{y_q} dy_1 \frac{\partial  \Delta \tilde{H}_{\vq,2}^{(n)}(x_q,y_1)}{\partial x_q}.
    \end{equation}
    It can be represented as $\mathcal{I} = (1/2)\mathcal{I} + (1/2) \mathcal{I}$. 
    In the second term, we perform the following substitutions sequentially: first, we replace $\phi \rightarrow \phi + \pi$      during the integration over  $\phi$
    and then we replace $y_1 = - y_1$ during the integration over $y_1$. 
    After applying these transformations, the two terms can be combined, yielding
    \begin{equation}
        \mathcal{I} = \frac{1}{2} \int\limits_0^{2\pi} d\phi \,  y_q \int\limits_{-\infty}^{\infty} dy_1 \frac{\partial  \Delta \tilde{H}_{\vq,2}^{(n)}(x_q,y_1)}{\partial x_q}.        
    \end{equation}
    Thus, instead of \eqref{eq:ap:F_H2}, we can write 
    \begin{equation}
        {F}_{\rm m \rightarrow v}^{(np)i} = 
            \frac{2}{(2\pi \hbar)^3} \int\limits_0^\infty dq\, q^2 
             \frac{\partial \mN_{\vq,0 }^{ (n)}}{\partial \varepsilon_\vq^{(n)}} 
             \int\limits_{4\pi} d\Omega_q\, q^i
             \int\limits_0^{2\pi} 
             d \phi \,
             y_q
             \left[ \frac{\vVrn \vq_\perp}{q_\perp} \Delta \tilde{H}_{\vq,2}^{(n)}(x_q,y_q) -  \vVrn \frac{\vec{e}_z \times \vq_\perp}{q_\perp}  \frac{1}{2} \int\limits_{-\infty}^{\infty} dy_1 \frac{\partial  \Delta \tilde{H}_{\vq,2}^{(n)}(x_q,y_1)}{\partial x_q} 
             \right].
    \end{equation}
    One can see that the first term in the square brackets vanishes upon integration over $\phi$, since it changes sign under the transformation $\phi \to \phi+\pi$. Likewise, the second term in the square brackets vanishes, because for any function $F(s)$ integrable on the interval $[-1,1]$, one has $\int_0^{2\pi} d\phi \cos\phi \, F(\sin\phi)= 0$.

\bibliography{mn-jour,paper}

@STRING{as = "Appl. Spectrosc."}

@STRING{aap = "A\&A"}

@STRING{apj = "ApJ"}

@STRING{mn = "MNRAS"}

@STRING{on = "Opt. News"}

@STRING{prc = "Phys. Rev. C"}

@STRING{prd = "Phys. Rev. D"}

@STRING{prl = "Phys. Rev. Lett."}

@ARTICLE{pcc19,
       author = {{Potekhin}, A.~Y. and {Chugunov}, A.~I. and {Chabrier}, G.},
        title = "{Thermal evolution and quiescent emission of transiently accreting neutron stars}",
      journal = aap,
         year = 2019,
        month = sep,
       volume = {629},
          eid = {A88},
        pages = {A88},
          doi = {10.1051/0004-6361/201936003},
archivePrefix = {arXiv},
       eprint = {1907.08299},
 primaryClass = {astro-ph.HE},
       adsurl = {https://ui.adsabs.harvard.edu/abs/2019A&A...629A..88P}
}

@ARTICLE{brown_etal18,
       author = {{Brown}, Edward F. and {Cumming}, Andrew and {Fattoyev}, Farrukh J. and {Horowitz}, C.~J. and {Page}, Dany and {Reddy}, Sanjay},
        title = "{Rapid Neutrino Cooling in the Neutron Star MXB 1659-29}",
      journal = prl,
         year = 2018,
        month = may,
       volume = {120},
       number = {18},
          eid = {182701},
        pages = {182701},
          doi = {10.1103/PhysRevLett.120.182701},
archivePrefix = {arXiv},
       eprint = {1801.00041},
 primaryClass = {astro-ph.HE},
       adsurl = {https://ui.adsabs.harvard.edu/abs/2018PhRvL.120r2701B}
}

@ARTICLE{ho11,
       author = {{Ho}, Wynn C.~G.},
        title = "{Superfluid effects on gauging core temperatures of neutron stars in low-mass X-ray binaries}",
      journal = mn,
         year = 2011,
        month = nov,
       volume = {418},
       number = {1},
        pages = {L99-L103},
          doi = {10.1111/j.1745-3933.2011.01152.x},
archivePrefix = {arXiv},
       eprint = {1109.0018},
 primaryClass = {astro-ph.HE},
       adsurl = {https://ui.adsabs.harvard.edu/abs/2011MNRAS.418L..99H}
}

@ARTICLE{wbs20,
       author = {{Wei}, Jin-Biao and {Burgio}, Fiorella and {Schulze}, Hans-Josef},
        title = "{Nuclear Pairing Gaps and Neutron Star Cooling}",
      journal = {Universe},
         year = 2020,
        month = aug,
       volume = {6},
       number = {8},
          eid = {115},
        pages = {115},
          doi = {10.3390/universe6080115},
archivePrefix = {arXiv},
       eprint = {2010.03916},
 primaryClass = {nucl-th},
       adsurl = {https://ui.adsabs.harvard.edu/abs/2020Univ....6..115W}
}

@ARTICLE{fortin_etal18,
       author = {{Fortin}, M. and {Taranto}, G. and {Burgio}, G.~F. and {Haensel}, P. and {Schulze}, H.-J. and {Zdunik}, J.~L.},
        title = "{Thermal states of neutron stars with a consistent model of interior}",
      journal = mn,
         year = 2018,
        month = apr,
       volume = {475},
       number = {4},
        pages = {5010-5022},
          doi = {10.1093/mnras/sty147},
archivePrefix = {arXiv},
       eprint = {1709.04855},
 primaryClass = {astro-ph.HE},
       adsurl = {https://ui.adsabs.harvard.edu/abs/2018MNRAS.475.5010F}
}

@ARTICLE{potekhin_etal20,
       author = {{Potekhin}, A.~Y. and {Zyuzin}, D.~A. and {Yakovlev}, D.~G. and {Beznogov}, M.~V. and {Shibanov}, Yu A.},
        title = "{Thermal luminosities of cooling neutron stars}",
      journal = mn,
         year = 2020,
        month = aug,
       volume = {496},
       number = {4},
        pages = {5052-5071},
          doi = {10.1093/mnras/staa1871},
archivePrefix = {arXiv},
       eprint = {2006.15004},
 primaryClass = {astro-ph.HE},
       adsurl = {https://ui.adsabs.harvard.edu/abs/2020MNRAS.496.5052P}
}

@ARTICLE{gck14,
       author = {{Gusakov}, Mikhail E. and {Chugunov}, Andrey I. and {Kantor}, Elena M.},
        title = "{Explaining observations of rapidly rotating neutron stars in low-mass x-ray binaries}",
      journal = prd,
         year = 2014,
        month = sep,
       volume = {90},
       number = {6},
          eid = {063001},
        pages = {063001},
          doi = {10.1103/PhysRevD.90.063001},
archivePrefix = {arXiv},
       eprint = {1305.3825},
 primaryClass = {astro-ph.SR},
       adsurl = {https://ui.adsabs.harvard.edu/abs/2014PhRvD..90f3001G}
}

@ARTICLE{moraga_etal25,
       author = {{Moraga}, N.~A. and {Castillo}, F. and {Ofengeim}, D.~D. and {Reisenegger}, A. and {Valdivia}, J.~A. and {Gusakov}, M.~E. and {Kantor}, E.~M. and {Potekhin}, A.~Y.},
        title = "{Magnetothermal evolution of neutron star cores in the weak-coupling regime: Implications of ambipolar diffusion for the quiescent x-ray luminosity of magnetars}",
      journal = {\prd},
         year = 2025,
        month = oct,
       volume = {112},
       number = {8},
          eid = {083022},
        pages = {083022},
          doi = {10.1103/16ny-kw3h},
archivePrefix = {arXiv},
       eprint = {2505.18733},
 primaryClass = {astro-ph.HE},
       adsurl = {https://ui.adsabs.harvard.edu/abs/2025PhRvD.112h3022M}
}

@ARTICLE{bl15,
       author = {{Beloborodov}, Andrei M. and {Li}, Xinyu},
        title = "{Magnetar Heating}",
      journal = apj,
         year = 2016,
        month = dec,
       volume = {833},
       number = {2},
          eid = {261},
        pages = {261},
          doi = {10.3847/1538-4357/833/2/261},
archivePrefix = {arXiv},
       eprint = {1605.09077},
 primaryClass = {astro-ph.HE},
       adsurl = {https://ui.adsabs.harvard.edu/abs/2016ApJ...833..261B}
}

@ARTICLE{kaminker_etal06,
       author = {{Kaminker}, A.~D. and {Yakovlev}, D.~G. and {Potekhin}, A.~Y. and {Shibazaki}, N. and {Shternin}, P.~S. and {Gnedin}, O.~Y.},
        title = "{Magnetars as cooling neutron stars with internal heating}",
      journal = mn,
         year = 2006,
        month = sep,
       volume = {371},
       number = {1},
        pages = {477-483},
          doi = {10.1111/j.1365-2966.2006.10680.x},
archivePrefix = {arXiv},
       eprint = {astro-ph/0605449},
 primaryClass = {astro-ph},
       adsurl = {https://ui.adsabs.harvard.edu/abs/2006MNRAS.371..477K}
}

@ARTICLE{kkpy14,
       author = {{Kaminker}, A.~D. and {Kaurov}, A.~A. and {Potekhin}, A.~Y. and {Yakovlev}, D.~G.},
        title = "{Thermal emission of neutron stars with internal heaters}",
      journal = mn,
         year = 2014,
        month = aug,
       volume = {442},
       number = {4},
        pages = {3484-3494},
          doi = {10.1093/mnras/stu1102},
archivePrefix = {arXiv},
       eprint = {1406.0723},
 primaryClass = {astro-ph.HE},
       adsurl = {https://ui.adsabs.harvard.edu/abs/2014MNRAS.442.3484K}
}

@ARTICLE{dg17,
       author = {{Dommes}, V.~A. and {Gusakov}, M.~E.},
        title = "{Vortex buoyancy in superfluid and superconducting neutron stars}",
      journal = mn,
         year = 2017,
        month = may,
       volume = {467},
       number = {1},
        pages = {L115-L119},
          doi = {10.1093/mnrasl/slx011},
archivePrefix = {arXiv},
       eprint = {1701.06870},
 primaryClass = {astro-ph.HE},
       adsurl = {https://ui.adsabs.harvard.edu/abs/2017MNRAS.467L.115D}
}

@ARTICLE{SedrakianClark2019,
       author = {{Sedrakian}, Armen and {Clark}, John W.},
        title = "{Superfluidity in nuclear systems and neutron stars}",
      journal = {European Physical Journal A},
         year = 2019,
        month = sep,
       volume = {55},
       number = {9},
          eid = {167},
        pages = {167},
          doi = {10.1140/epja/i2019-12863-6},
archivePrefix = {arXiv},
       eprint = {1802.00017},
 primaryClass = {nucl-th},
       adsurl = {https://ui.adsabs.harvard.edu/abs/2019EPJA...55..167S}
}

@ARTICLE{GusakovHaensel2005,
   author = {{Gusakov}, M.~E. and {Haensel}, P.},
    title = "{The entrainment matrix of a superfluid neutron proton mixture at a finite temperature}",
  journal = {Nuclear Physics A},
   eprint = {arXiv:astro-ph/0508104},
     year = 2005,
    month = nov,
   volume = 761,
    pages = {333-348},
      doi = {10.1016/j.nuclphysa.2005.07.005}
}

@ARTICLE{Jones1991b,
       author = {{Jones}, P.~B.},
        title = "{Neutron superfluid spin-down and magnetic field decay in pulsars.}",
      journal = mn,
         year = 1991,
        month = nov,
       volume = {253},
        pages = {279},
          doi = {10.1093/mnras/253.2.279}
}

@ARTICLE{Jones2006,
       author = {{Jones}, P.~B.},
        title = "{Type II superconductivity and magnetic flux transport in neutron stars}",
      journal = mn,
         year = 2006,
        month = jan,
       volume = {365},
       number = {1},
        pages = {339-344},
          doi = {10.1111/j.1365-2966.2005.09724.x},
archivePrefix = {arXiv},
       eprint = {astro-ph/0510396},
 primaryClass = {astro-ph},
}

@ARTICLE{Jones2009,
       author = {{Jones}, P.~B.},
        title = "{Fermion zero-mode influence on neutron-star magnetic field evolution}",
      journal = mn,
         year = 2009,
        month = aug,
       volume = {397},
       number = {2},
        pages = {1027-1031},
          doi = {10.1111/j.1365-2966.2009.15016.x},
archivePrefix = {arXiv},
       eprint = {0902.0478},
 primaryClass = {astro-ph.HE}
}

@ARTICLE{AlfordSedrakian2010,
       author = {{Alford}, Mark G. and {Sedrakian}, Armen},
        title = "{Color-magnetic flux tubes in quark matter cores of neutron stars}",
      journal = {Journal of Physics G Nuclear Physics},
         year = 2010,
        month = jul,
       volume = {37},
       number = {7},
          eid = {075202},
        pages = {075202},
          doi = {10.1088/0954-3899/37/7/075202},
archivePrefix = {arXiv},
       eprint = {1001.3346},
 primaryClass = {astro-ph.SR}
}

@ARTICLE{BransgroveLevinBeloborodov2018,
       author = {{Bransgrove}, Ashley and {Levin}, Yuri and {Beloborodov}, Andrei},
        title = "{Magnetic field evolution of neutron stars - I. Basic formalism, numerical techniques and first results}",
      journal = mn,
         year = 2018,
        month = jan,
       volume = {473},
       number = {2},
        pages = {2771-2790},
          doi = {10.1093/mnras/stx2508},
archivePrefix = {arXiv},
       eprint = {1709.09167},
 primaryClass = {astro-ph.HE}
}

@ARTICLE{GraberEtAl2015,
       author = {{Graber}, Vanessa and {Andersson}, Nils and {Glampedakis}, Kostas and {Lander}, Samuel K.},
        title = "{Magnetic field evolution in superconducting neutron stars}",
      journal = mn,
         year = 2015,
        month = oct,
       volume = {453},
       number = {1},
        pages = {671-681},
          doi = {10.1093/mnras/stv1648},
archivePrefix = {arXiv},
       eprint = {1505.00124},
 primaryClass = {astro-ph.SR}
}

@ARTICLE{GlampedakisAnderssonSamuelsson2011,
   author = {{Glampedakis}, K. and {Andersson}, N. and {Samuelsson}, L.},
    title = "{Magnetohydrodynamics of superfluid and superconducting neutron star cores}",
  journal = mn,
archivePrefix = "arXiv",
   eprint = {1001.4046},
 primaryClass = "astro-ph.SR",
     year = 2011,
    month = jan,
   volume = 410,
    pages = {805-829},
      doi = {10.1111/j.1365-2966.2010.17484.x},
}

@ARTICLE{AlparLangerSauls1984,
   author = {{Alpar}, M.~A. and {Langer}, S.~A. and {Sauls}, J.~A.},
    title = "{Rapid postglitch spin-up of the superfluid core in pulsars}",
  journal = apj,
     year = 1984,
    month = jul,
   volume = 282,
    pages = {533-541},
    doi = {10.1086/162232}
}

@ARTICLE{Gusakov2010,
       author = {{Gusakov}, Mikhail E.},
        title = "{Transport equations and linear response of superfluid Fermi mixtures in neutron stars}",
      journal = {\prc},
         year = 2010,
        month = feb,
       volume = {81},
       number = {2},
          eid = {025804},
        pages = {025804},
          doi = {10.1103/PhysRevC.81.025804},
archivePrefix = {arXiv},
       eprint = {1001.4452},
 primaryClass = {astro-ph.SR},
}

@ARTICLE{ChamelHaensel2006,
       author = {{Chamel}, Nicolas and {Haensel}, Pawel},
        title = "{Entrainment parameters in a cold superfluid neutron star core}",
      journal = {\prc},
         year = 2006,
        month = apr,
       volume = {73},
       number = {4},
          eid = {045802},
        pages = {045802},
          doi = {10.1103/PhysRevC.73.045802},
archivePrefix = {arXiv},
       eprint = {nucl-th/0603018},
 primaryClass = {nucl-th}
}

@ARTICLE{AllardChamel2021,
       author = {{Allard}, V. and {Chamel}, N.},
        title = "{Entrainment effects in neutron-proton mixtures within the nuclear energy-density functional theory. II. Finite temperatures and arbitrary currents}",
      journal = {\prc},
         year = 2021,
        month = feb,
       volume = {103},
       number = {2},
          eid = {025804},
        pages = {025804},
          doi = {10.1103/PhysRevC.103.025804},
archivePrefix = {arXiv},
       eprint = {2006.15317},
 primaryClass = {nucl-th}
}

@ARTICLE{Leggett1965,
       author = {{Leggett}, A.~J.},
        title = "{Theory of a Superfluid Fermi Liquid. I. General Formalism and Static Properties}",
      journal = {Physical Review},
         year = 1965,
        month = dec,
       volume = {140},
       number = {6A},
        pages = {1869-1888},
          doi = {10.1103/PhysRev.140.A1869}
}

@ARTICLE{AronovEtAl1981,
       author = {{Aronov}, A.~G. and {Gal'Perin}, Yu. M. and {Gurevich}, V.~L. and {Kozub}, V.~I.},
        title = "{The Boltzmann-equation description of transport in superconductors}",
      journal = {Advances in Physics},
         year = 1981,
        month = jul,
       volume = {30},
       number = {4},
        pages = {539-592},
          doi = {10.1080/00018738100101407}
}

@book{PinesNozieres,
  title={The Theory of Quantum Liquids: Normal Fermi liquids},
  author={Pines, D. and Nozi{\`e}res, P.},
  lccn={66013108},
  series={Advanced book classics series},
  year={1966},
  publisher={W.A. Benjamin},
  address = {New York}
}

@ARTICLE{GG23,
       author = {{Goglichidze}, O.~A. and {Gusakov}, M.~E.},
        title = "{Diffusion in superfluid Fermi mixtures: General formalism}",
      journal = {\prc},
         year = 2023,
        month = aug,
       volume = {108},
       number = {2},
          eid = {025814},
        pages = {025814},
          doi = {10.1103/PhysRevC.108.025814},
archivePrefix = {arXiv},
       eprint = {2309.03313},
 primaryClass = {astro-ph.HE}
}

@ARTICLE{GG25,
       author = {{Goglichidze}, Oleg A. and {Gusakov}, Mikhail E.},
        title = "{Nonpurely transverse Magnus force in superconducting neutron stars}",
      journal = {\prd},
         year = 2025,
        month = may,
       volume = {111},
       number = {10},
          eid = {103051},
        pages = {103051},
          doi = {10.1103/1l6g-t95k},
archivePrefix = {arXiv},
       eprint = {2505.05628},
 primaryClass = {astro-ph.HE},
       adsurl = {https://ui.adsabs.harvard.edu/abs/2025PhRvD.111j3051G}
}

@book{SoninBook,
  title={Dynamics of Quantised Vortices in Superfluids},
  author={Sonin, E.B.},
  isbn={9781316455906},
  year={2016},
  publisher={Cambridge University Press},
   address = {Cambridge}
}

@ARTICLE{Gusakov2019,
       author = {{Gusakov}, M.~E.},
        title = "{Force on proton vortices in superfluid neutron stars}",
      journal = mn,
         year = 2019,
        month = jun,
       volume = {485},
       number = {4},
        pages = {4936-4950},
          doi = {10.1093/mnras/stz657},
archivePrefix = {arXiv},
       eprint = {1904.01363},
 primaryClass = {astro-ph.HE}
}

@book{deGennes_book,
  title={Superconductivity of Metals and Alloys},
  author={de Gennes, P.G.},
  lccn={lc65017013},
  series={Frontiers in physics},
  year={1966},
  publisher={W.A. Benjamin},
  address = {New York}
}

@book{LL2,
  title={The Classical Theory of Fields: Volume 2},
  author={Landau, L.D.   and Lifshits, E.M.},
  isbn={9780750627689},
  lccn={75004737},
  series={Course of theoretical physics},
  year={1975},
  publisher={Elsevier Science},
  address = {Oxford}
}

@ARTICLE{KopninKravtsov1976b,
       author = {{Kopnin}, N.~B. and {Kravtsov}, V.~E.},
        title = "{Conductivity and Hall effect of pure type-II superconductors at low temperatures}",
      journal = {Soviet Journal of Experimental and Theoretical Physics Letters},
         year = 1976,
        month = jun,
       volume = {23},
        pages = {578}
}

@ARTICLE{Sonin1975,
       author = {{Sonin}, {\'E}. B.},
        title = "{Friction between the normal component and vortices in rotating superfluid helium}",
      journal = {Soviet Journal of Experimental and Theoretical Physics},
         year = 1975,
        month = sep,
       volume = {42},
        pages = {469}
}

@ARTICLE{SourieChamel2020,
       author = {{Sourie}, Aur{\'e}lien and {Chamel}, Nicolas},
        title = "{Force on a neutron quantized vortex pinned to proton fluxoids in the superfluid core of cold neutron stars}",
      journal = mn,
         year = 2020,
        month = mar,
       volume = {493},
       number = {1},
        pages = {382-389},
          doi = {10.1093/mnras/staa253},
archivePrefix = {arXiv},
       eprint = {2001.08964},
 primaryClass = {astro-ph.HE}
}

@ARTICLE{Kopnin1995,
       author = {{Kopnin}, N.~B.},
        title = "{Theory of mutual friction in superfluid $^{3}$He at low temperatures}",
      journal = {Physica B Condensed Matter},
         year = 1995,
        month = feb,
       volume = {210},
       number = {3-4},
        pages = {267-286},
          doi = {10.1016/0921-4526(94)01113-F}
}

@ARTICLE{PearsonEtAl2019,
       author = {{Pearson}, J.~M. and {Chamel}, N. and {Potekhin}, A.~Y. and {Fantina}, A.~F. and {Ducoin}, C. and {Dutta}, A.~K. and {Goriely}, S.},
        title = "{Unified equations of state for cold non-accreting neutron stars with Brussels-Montreal functionals - I. Role of symmetry energy}",
      journal = mn,
         year = 2018,
        month = dec,
       volume = {481},
       number = {3},
        pages = {2994-3026},
          doi = {10.1093/mnras/sty2413},
archivePrefix = {arXiv},
       eprint = {1903.04981},
 primaryClass = {astro-ph.HE}
}

@ARTICLE{GorielyChamelPearson2013b,
       author = {{Goriely}, S. and {Chamel}, N. and {Pearson}, J.~M.},
        title = "{Further explorations of Skyrme-Hartree-Fock-Bogoliubov mass formulas. XIII. The 2012 atomic mass evaluation and the symmetry coefficient}",
      journal = {\prc},
         year = 2013,
        month = aug,
       volume = {88},
       number = {2},
          eid = {024308},
        pages = {024308},
          doi = {10.1103/PhysRevC.88.024308}
}

\end{document}